\documentclass[twocolumn,trackchanges]{aastex63}

\usepackage{amsmath}
\usepackage{multirow}

\shorttitle{Find SB using LAMOST MRS Data}
\shortauthors{LI et al.}
\date{December 2023}

\begin{document}

\title{Mining double-line spectroscopic candidates in the LAMOST medium-resolution spectroscopic survey using human-AI hybrid method}

\author[0000-0002-4459-1848]{Shan-shan Li}
\affiliation{National Astronomical Data Center of China, National Astronomical Observatories, Chinese Academy of Sciences, Beijing 100101, China}
\affiliation{School of Astronomy and Space Science, University of Chinese Academy of Sciences, Beijing 100049, China}

\author[0000-0002-6647-3957]{Chun-qian Li}
\affiliation{CAS Key Laboratory of Optical Astronomy, National Astronomical Observatories, Chinese Academy of Sciences, Beijing 100101, China}
\affiliation{School of Astronomy and Space Science, University of Chinese Academy of Sciences, Beijing 100049, China}

\author{Chang-hua Li}
\affiliation{National Astronomical Data Center of China, National Astronomical Observatories, Chinese Academy of Sciences, Beijing 100101, China}

\author[0000-0002-8669-5370]{Dong-wei Fan}
\affiliation{National Astronomical Data Center of China, National Astronomical Observatories, Chinese Academy of Sciences, Beijing 100101, China}

\author[0000-0002-7397-811X]{Yun-fei Xu}
\affiliation{National Astronomical Data Center of China, National Astronomical Observatories, Chinese Academy of Sciences, Beijing 100101, China}

\author{Lin-ying Mi}
\affiliation{National Astronomical Data Center of China, National Astronomical Observatories, Chinese Academy of Sciences, Beijing 100101, China}
\affiliation{School of Astronomy and Space Science, University of Chinese Academy of Sciences, Beijing 100049, China}

\author[0000-0002-7456-1826]{Chen-zhou Cui}
\affiliation{National Astronomical Data Center of China, National Astronomical Observatories, Chinese Academy of Sciences, Beijing 100101, China}
\affiliation{School of Astronomy and Space Science, University of Chinese Academy of Sciences, Beijing 100049, China}
\email{ccz@nao.cas.cn}

\author[0000-0002-0349-7839]{Jian-rong Shi}
\affiliation{CAS Key Laboratory of Optical Astronomy, National Astronomical Observatories, Chinese Academy of Sciences, Beijing 100101, China}
\affiliation{School of Astronomy and Space Science, University of Chinese Academy of Sciences, Beijing 100049, China}
\email{sjr@nao.cas.cn}

\begin{abstract}

We utilize a hybrid approach that integrates the traditional cross-correlation function (CCF) and machine learning to detect spectroscopic multi-systems, specifically focusing on double-line spectroscopic binary (SB2). Based on the ninth data release (DR9) of the Large Sky Area Multi-Object Fiber Spectroscopic Telescope (LAMOST), which includes a medium-resolution survey (MRS) containing 29,920,588 spectra, we identify 27,164 double-line and 3124 triple-line spectra, corresponding to 7096 SB2 candidates and 1903 triple-line spectroscopic binary (SB3) candidates, respectively, representing about 1\% of the selection dataset from LAMOST-MRS DR9. Notably, 70.1\% of the SB2 candidates and 89.6\% of the SB3 candidates are newly identified. Compared to using only the traditional CCF technique, our method significantly improves the efficiency of detecting SB2, saves time on visual inspections by a factor of four.

\end{abstract}


\section{Introduction}

Multiple star systems, also called multi-systems, including binary star systems, are common and important celestial bodies in astronomical studies. Approximately half of the stars in the Milky Way reside in multiple star systems \citep{Duchene_2013, Raghavan_2010}. Studying multiple star systems provides insights into the processes of stellar formation and evolution, wherein the details of material exchange and interactions clearly govern the evolutionary paths and final outcomes \citep{Sana_2011, Han_2020}. Additionally, binary star systems serve as the only known direct method to determine accurate stellar masses \citep{Southworth_2020}. As for high-energy astrophysics and frontier research fields, such as the multi-band and time-domain astronomy studies, a more complete and uniform sample set of multi-systems would help to constrain theoretical models.

Due to the Doppler shift resulting from the differing radial velocities (RVs) of each component in a binary star system, we can measure the splitting of spectral lines. Furthermore, if the binary stars have similar spectral types but significant differences in their RVs, they are more likely to be observed as double-lined spectroscopic binaries. Spectroscopic methods are used to search for such spectroscopic binaries (SBs), and the systems discovered are called double-line, triple-line, quadruple-line spectroscopic binaries and so on (SB\textit{n}s; $n>1$) depending on the split number of spectral lines. There are also single-line spectroscopic binaries (SB1s) in which only the spectrum of one star can be clearly observed, often distinguished by variations in RV.

First published in 2004, the Ninth Catalog of Spectroscopic Binary Orbits ($\rm S_{B^9}$) is one of the most popular and influential catalogs of SBs, including over 4000 SBs, about one-third of them are SB2s \citep[][and the latest online version of the SB9 catalogue]{Pourbaix_2004}. Recent large spectroscopic sky surveys make it possible to systematically find SBs in large quantities. Many works searching for SBs have utilized the survey data from the Apache Point Observatory Galaxy Evolution Experiment \citep[APOGEE,][]{Prieto_2008}, which is part of the Sloan Digital Sky Survey (SDSS) \citep{York_2000}, including \citet{Fernandez_2017}, \citet{ElBadry_2018} and \citet{Whelan_2018, Whelan_2020}. Moreover, a pipeline has been developed to autonomously find SB2s from the APOGEE data, and over 7000 SB2s have been identified from DR16 \citep{Kounkel_2021}. From the Gaia-ESO survey \citep{Gilmore_2022, Randich_2022}, 641 SB1s, 342 SB2s, and 11 SB3s have been identified \citep{Merle_2017, Merle_2020}. Additionally, a recent study employing a new technique reports the identification of 322 SB2s, 10 SB3s, and 2 SB4s from a sample of 37,565 objects \citep{Swaelmen_2023}. Several studies on SBs have also been conducted based on the Radial Velocity Experiment \citep[RAVE,][]{Steinmetz_2006}, such as \citet{Matijevic_2010} and \citet{Birko_2019}. The Galactic Archaeology with HERMES survey \citep[GALAH,][]{Silva_2015} has also contributed to this area, as demonstrated by \citet{Traven_2020}.

The Large Sky Area Multi-Object Fiber Spectroscopic Telescope \citep[LAMOST,][]{Cui_2012}, also known as the Guo Shou Jing Telescope, is a special reflecting Schmidt telescope that enables a large spectroscopic sky survey with as many as 4000 optical fibers. It provides an ideal data source for SBs mining. The first stage of the LAMOST spectroscopic survey was from October 2011 to July 2017. In 2019, more than 250,000 spectroscopic binary or variable star candidates were discovered using this data \citep{Qian_binary_2019}. \citet{Tian_2020} utilized data from LAMOST Data Release 4 (DR4) to construct a catalog that includes approximately 60,000 binary star system candidates. The LAMOST Medium-Resolution Spectroscopic Survey (LAMOST-MRS), which began in September 2018 \citep{liu_lamost_2020}, has a resolving power of $\rm R\sim7500 $. \citet{li_double_2021} identified 3133 SB2 and 132 SB3 candidates using LAMOST-MRS DR7 data.  \citet{zhang_spectroscopic_2022} used LAMOST-MRS DR8 data to discover 2198 SB2 candidates, and \citet{Kovalev_2022} detected 2460  SB2 candidates.

In this paper, we use an innovative human and artificial intelligence (human-AI) hybrid method to measure RVs from LAMOST-MRS data and search for SB candidates. We use the data from the ninth data release (DR9) of LAMOST-MRS \footnote{https://www.lamost.org/dr9/v1.0}. As shown in Figure~\ref{fig_fig1SNR}, the LAMOST-MRS data includes a sufficient number of high signal-to-noise ratio (S/N) spectra. Each exposure of LAMOST-MRS spectrum includes a blue arm ($4950\sim5350$ \AA) and a red arm ($6300\sim6800$ \AA). We use the blue arm of each spectrum to calculate the cross-correlation function (CCF) and RVs because it contains more absorption lines. Based on the number of RVs, the spectra can be identified and classified as SBs or not, but the precision is relatively low. Therefore, visual inspection is needed, which is inefficient. We use machine learning (ML) methods to develop automatic classifiers that can replace human inspection, greatly improving the efficiency of searching for SBs using LAMOST-MRS data.

In Section 2, we focus on the data selection. Section 3 describes the steps and details of our method, including conventional CCF technique and the construction of ML classifiers. In Section 4, we present the results of the data processing and all SB2, SB3 candidates identified. The discussion and conclusion are presented in Sections 5 and 6, respectively.

\section{Data Selection}

\begin{figure*}
    \centering
	\begin{minipage}{0.49\linewidth}
	    \centering
		\includegraphics[width=0.95\linewidth]{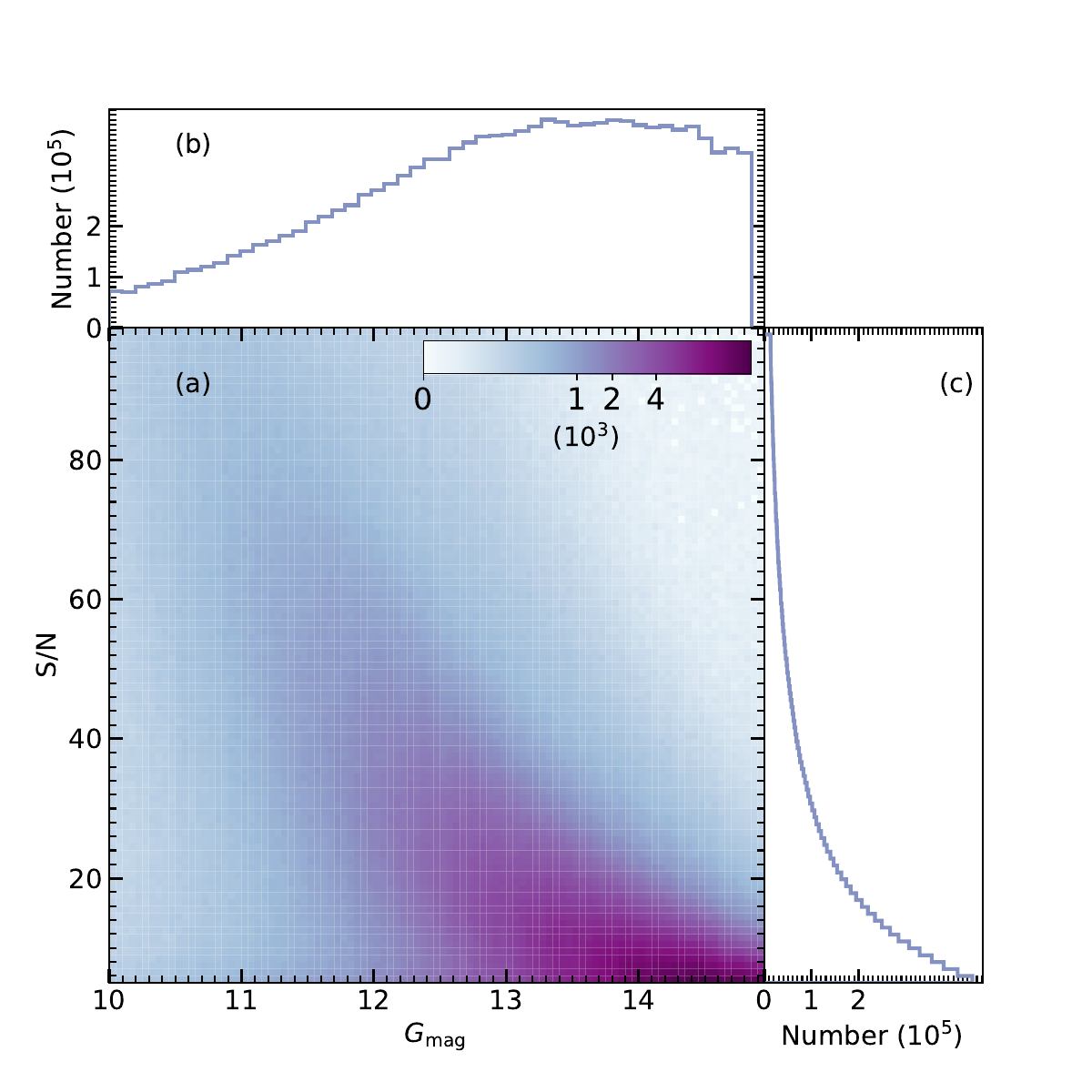}
	\end{minipage}
	\begin{minipage}{0.49\linewidth}
	    \centering
		\includegraphics[width=0.95\linewidth]{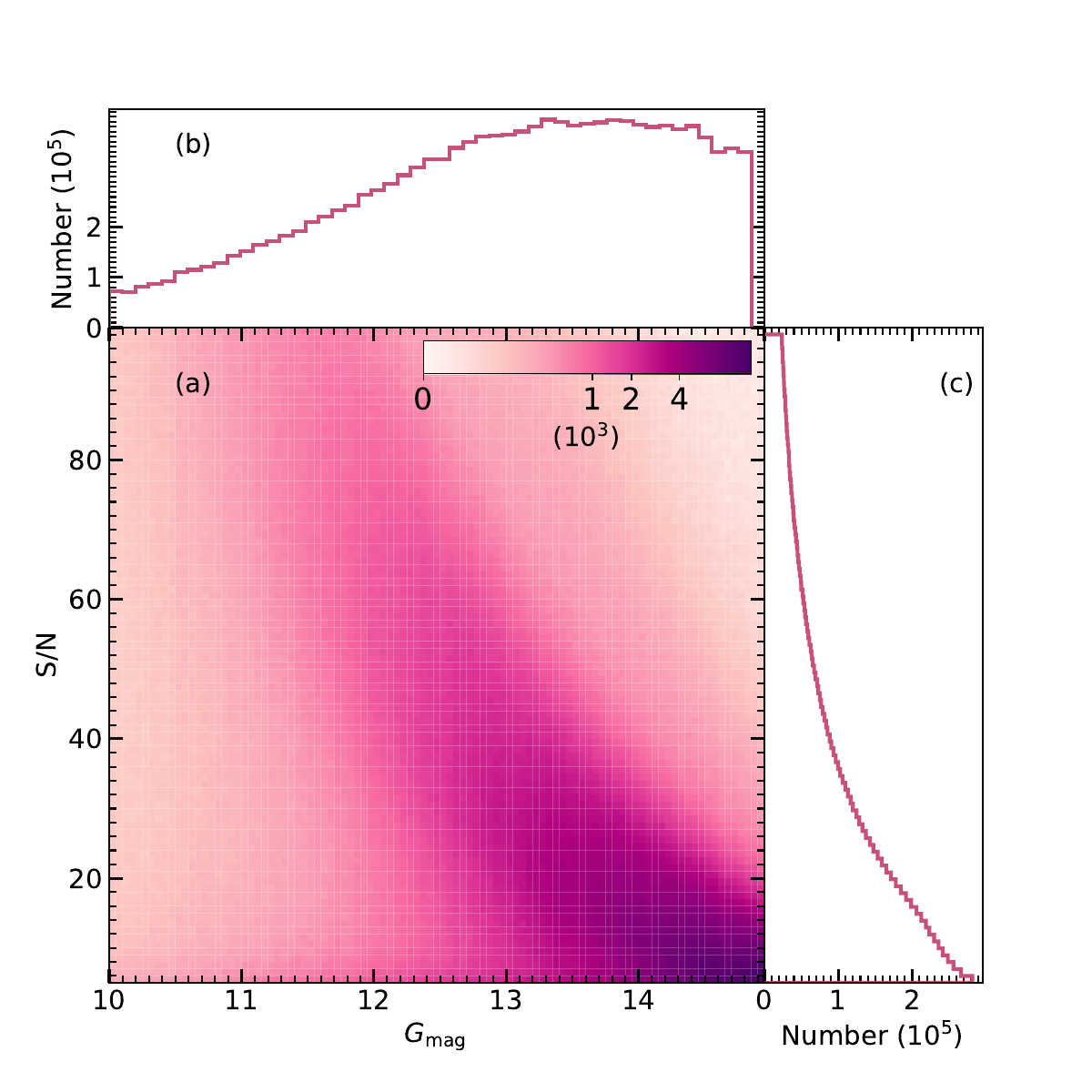}
	\end{minipage}
    \caption{The left panel displays the spectra from the blue arm of the LAMOST-MRS DR9 data, while the right panel presents the spectra from the red arm. In each panel, the parts labeled as a, b, and c represent the distribution of S/N versus G magnitude, the distribution of G magnitude, and the distribution of S/N, respectively.}
    \label{fig_fig1SNR}
\end{figure*}

The LAMOST-MRS DR9, released in March 2022, contains 29,920,588 spectra. The dataset includes all observational data newly processed by the latest pipeline, called LAMOST Stellar Parameter Pipeline \citep[LASP;][]{Luo_LAMOST_2015}. LAMOST-MRS DR9 includes both time-domain and non-time-domain surveys. In the time-domain survey, each target is observed multiple times over several nights, capturing both single-exposure spectra and coadded spectra (combined within one night). We focus exclusively on single-exposure data, indicated by a ``0'' in the \textit{coadd} field of the catalog. Additionally, each exposure in LAMOST-MRS yields two spectra: one from the blue arm and another one from the red arm. We select the blue arm spectra from the MRS data due to their abundance of absorption lines, indicated by a \textit{band} field value of ``B'' in the corresponding catalog files. Figure \ref{fig_fig1SNR} displays the distribution of S/N versus \textit{Gaia} G magnitude for LAMOST-MRS DR9 within the magnitude range of 10 to 15. Selecting spectra based on S/N is crucial for the search results. We choose spectra with S/N greater than 5 to fully exploit the potential of our method and maximize the discovery of SB2s and SB3s. Furthermore, we need to exclude fibers with issues caused by device malfunctions and problematic blue arm spectra. These issues are indicated in the catalog by the parameters \textit{fibermask} and \textit{band\_b} with ``0'' indicating no issues and ``1'' indicating problems. Following these criteria, as shown in Table 1, we have obtained a final sample of 6,565,721 spectra from 930,783 stars. For more information, please refer to the Data Release document of LAMOST-MRS DR9\footnote{https://www.lamost.org/dr9/v1.0/doc/mr-data-production-description}.

\begin{table}[ht]
    \tablenum{1}
    \label{tab:tableCriteria}
    \begin{center}
        \caption{The criteria we used for screening LAMOST-MRS DR9 data.}
        \begin{tabular}{ccc}
        \hline
        \hline
        Field & Criteria & Notes\\
        \hline
        band & B & Only blue arm spectrum\\
        coadd & 0 & No combined spectrum\\
        fibermask & 0 & Delete spectra from bad fiber\\
        $\rm bad\_b $ & 0 & Delete bad blue arm spectra \\
        S/N & $\rm \geq 5$ & \\
        \hline
        \end{tabular}
    \end{center}
\end{table}

\section{The Method}

Astronomical research has entered the era of big data, characterized by a rapid increase in data volume, with observational data nearly doubling every 16 months \citep{Zhang_2015}. This massive increase in astronomical data presents challenges for research, potentially causing important discoveries to be missed or delayed \citep{Smith_2023}. Meanwhile, AI and ML methods have shown a certain degree of effectiveness in source detection and classification \citep{Skoda_2020}, particularly for searching SB spectra in large spectroscopic sky survey data \citep{Traven_2020, li_double_2021, zhang_spectroscopic_2022}.  

Notably, in large spectroscopic sky surveys like LAMOST-MRS, spectra from binary star systems account for only a small portion of the total spectra \citep[1.2\% as shown by][]{li_double_2021}. This imbalance makes it very challenging to construct a training dataset with a similar distribution of binary star systems and non-binary star systems, which can significantly affect the results of classification \citep{He_2009}. Therefore, we use the traditional CCF technique for initial screening to detect SB2s and SB3s, which helps to increase the balance of the samples and improve the classification performance of the machine learning model. Additionally, the CCF technique is needed to measure the RV values of the SB candidates.

To maximize the advantages of conventional and ML methods, our study is mainly divided into two steps. First, we employ the CCF technique to calculate RVs using the observed and template spectra. This process results in a categorized list of spectra and the corresponding CCF data for each spectrum. Subsequently, multiple ML classifiers and an ensemble learning strategy are applied to the CCF data from the first step to identify SB2 and SB3 systems.

\subsection{Cross-Correlation Function} \label{section: CCF method}
The CCF technique is a conventional approach for detecting RV components and calculating RV values from spectra.

\subsubsection{CCF calculation} \label{section: CCF calculation}
We calculated the CCF based on the observed and template spectra with the classic normalized CCF calculation shown in Equation \ref{eq1} \citep{gubner2006probability,Zverko_2007},

\begin{equation}
\label{eq1}
CCF(v) = \sum_{i=1}^{n} (\frac{O_{i}-\overline{O}}{\sigma_{\rm O}})(\frac{T_{i,v}-\overline{T}}{\sigma_{\rm T}})
\end{equation}

\noindent here, \textit{O} stands for observation, while $T$ stands for template. $ O_{i} $ is the normalized flux of the spectrum at the wavelength-sampling point \textit{i}, $\overline{O} $ and $\sigma_{\rm O} $ are the mean flux and scatter of flux values, respectively. $T_{i,v}$ represents the normalized flux of the template spectrum at the wavelength sampling point $i$ under a Doppler shift corresponding to the velocity $v$. $\sigma_{\rm T} $ represents the standard deviation calculated from the total flux values of the template spectrum, and $\overline{T}$ is the arithmetic mean of the flux values of the template spectrum. The normalized CCF ranges from $-$1 to +1, with a fully correlated CCF at +1 and a fully anti-correlated CCF at $-$1. The normalized CCF allows for better detection of RV components and facilitates subsequent machine learning model training. Additionally, the CCF is calculated within the range of RV from -500 to +500 km/s with a step of 1 km/s.

We generate three spectral template using the stellar spectral synthesis program SPECTRUM \citep{Gray_1994} and ATLAS stellar atmospheric models \citep{Castelli_2003} under the assumptions of 1D local thermodynamic equilibrium (LTE). These template spectra include three stellar parameters: hot dwarfs ($T_{\rm eff}$  = 8000\,K, $\log{g}$ = 4.0\,dex, [Fe/H] = 0.0\,dex), cool dwarfs ($T_{\rm eff}$ = 5000\,K, $\log{g}$ = 4.0\,dex, [Fe/H] = 0.0\,dex) and cool giants ($T_{\rm eff}$  = 5000\,K, $\log{g}$ = 2.0\,dex, [Fe/H] = 0.0\,dex). The wavelength range and resolution of the template spectra match those of the LAMOST-MRS. After calculating the CCFs, the highest of the three CCFs is chosen for the next step in the multi-line spectral detection process.

\subsubsection{Detection of RV components} \label{section: RV components detection}

We adopted the method proposed by \citet{Merle_2017} to identify peaks in the CCF, corresponding to RV components in the spectra of SB. This semi-automatic process computes the first three derivatives of the CCF for a given spectrum and locates the peak positions, including blended peaks. The RV values of these components can be measured by the positions where the third derivative of CCF crosses zero during the ascending phase.

\begin{figure*}
	\centering
	\begin{minipage}{0.49\linewidth}
	    \centering
		\includegraphics[width=0.79\linewidth]{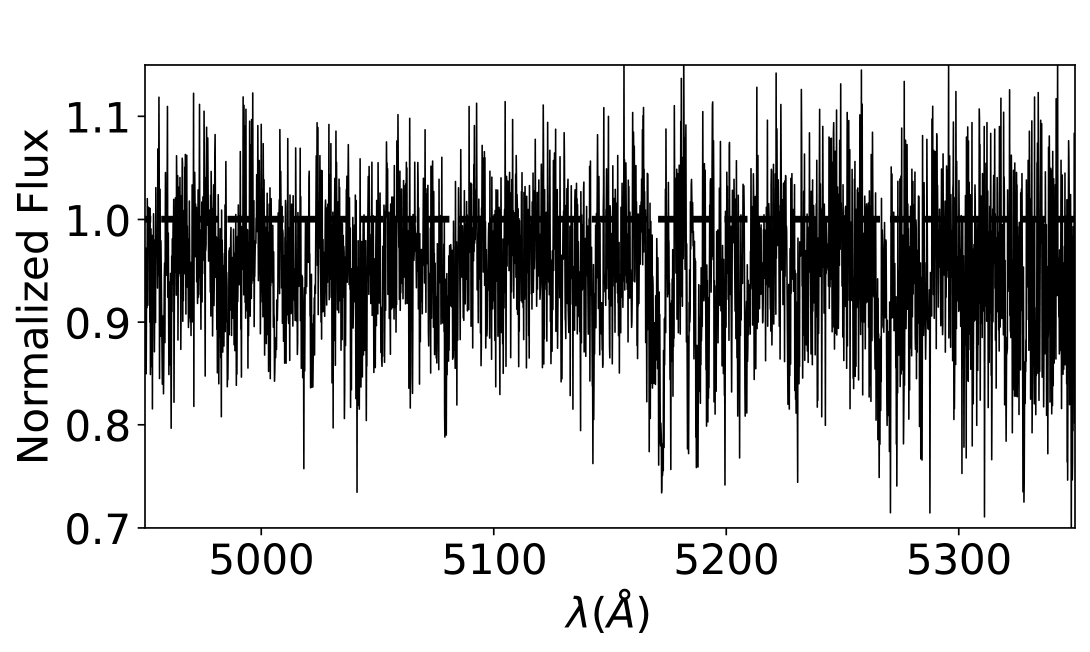}
	\end{minipage}
	\begin{minipage}{0.49\linewidth}
	    \centering
		\includegraphics[width=0.78\linewidth]{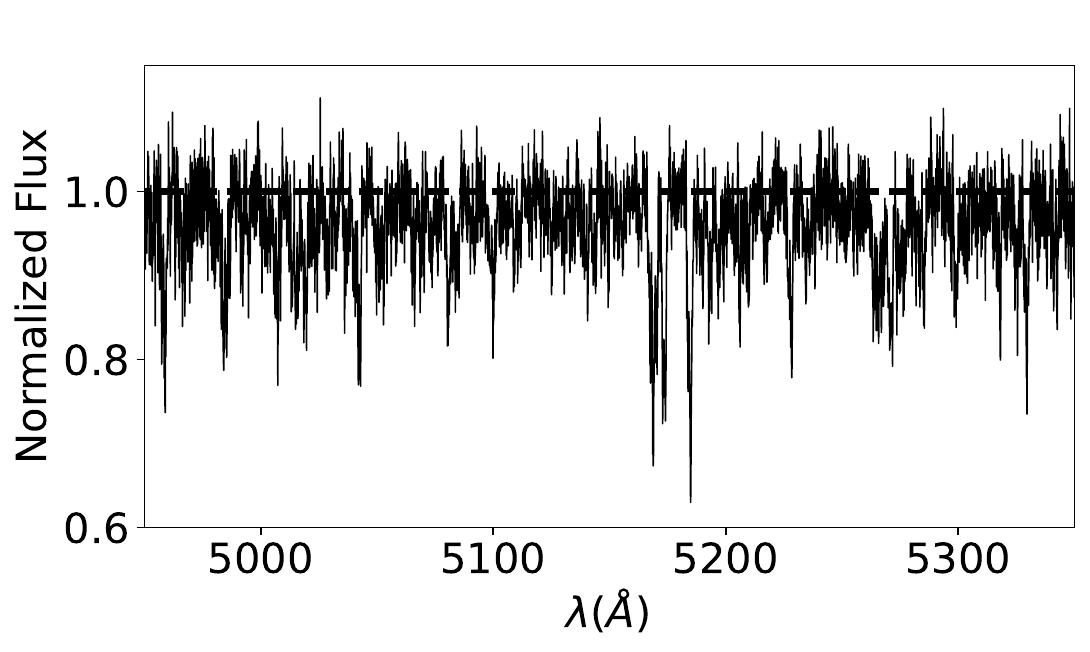}
	\end{minipage}
	\begin{minipage}{0.49\linewidth}
	    \centering
		\includegraphics[width=0.85\linewidth]{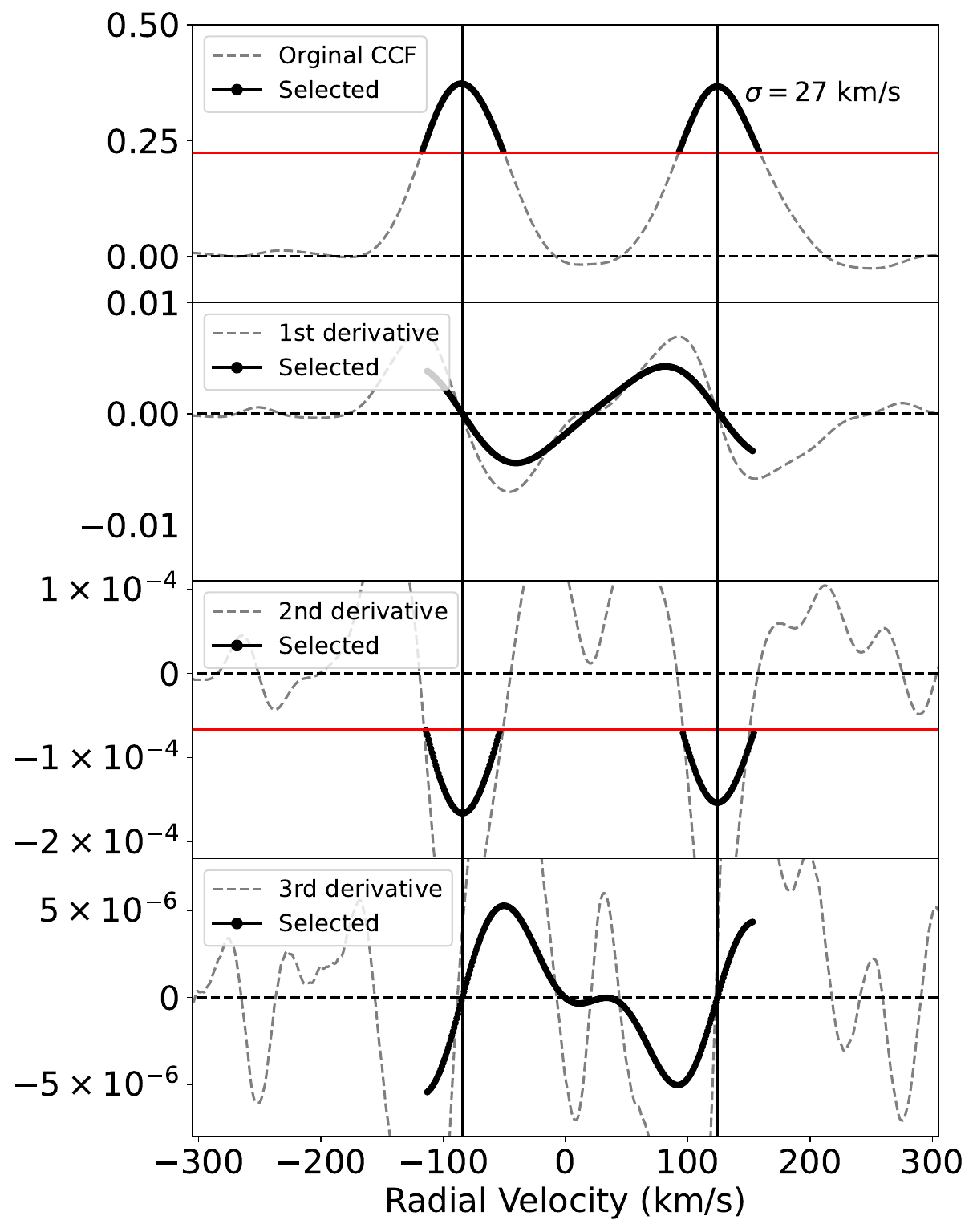}
	\end{minipage}
	\begin{minipage}{0.49\linewidth}
	    \centering
		\includegraphics[width=0.85\linewidth]{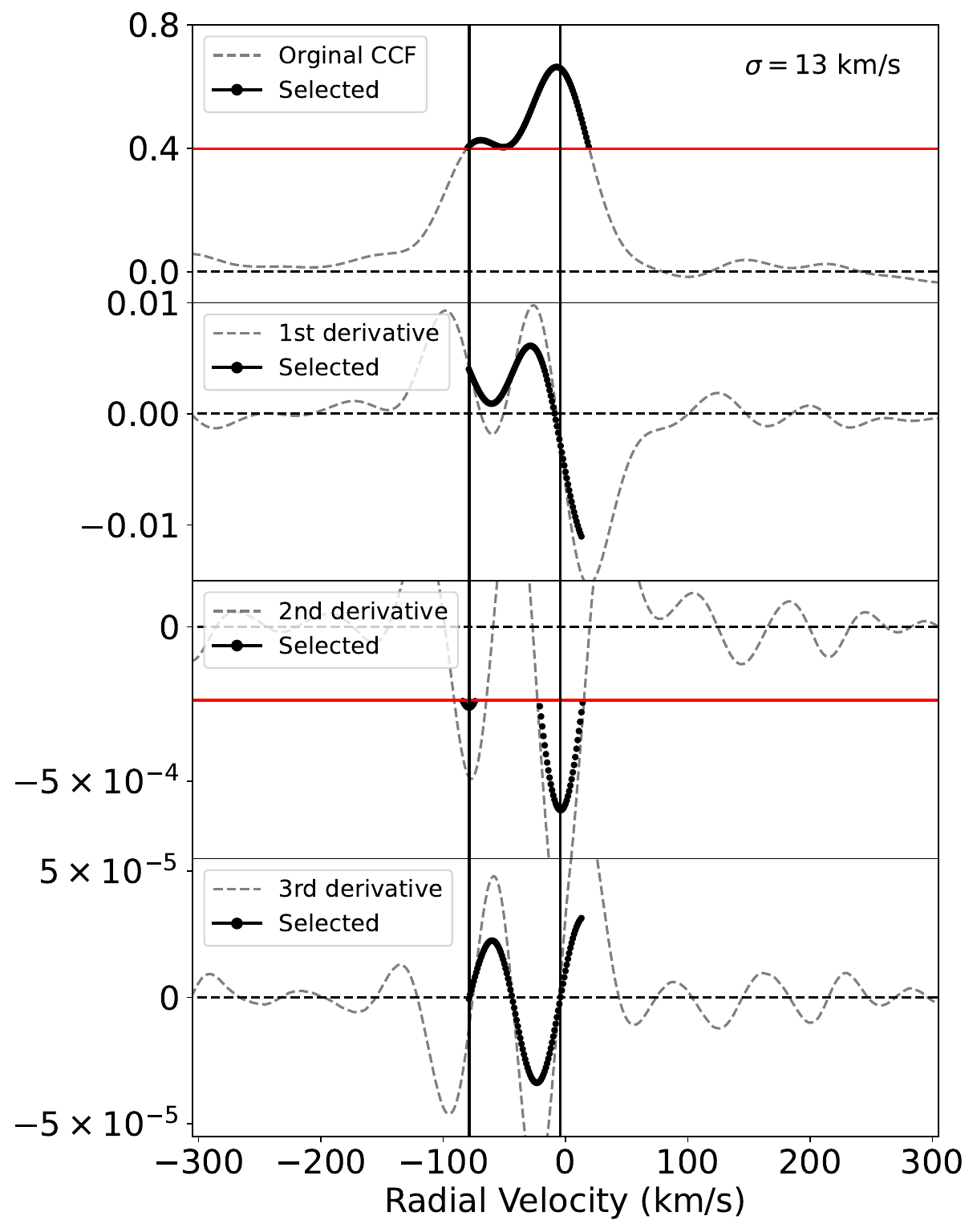}
	\end{minipage}
    \caption{The normalized spectra, CCFs, and derivatives of two SB2 candidates. In the left panel, the final $\sigma$ of the system is 27 km/s. In the right panel, the CCF exhibits significant peak blending; the first derivative cannot distinguish the peaks well, but they can be identified using the third derivative. Black solid lines are used to draw the selected range of smoothed CCFs and derivatives, while gray dashed lines illustrate the original CCFs. Red horizontal lines indicate the thresholds (above 60\% for CCF values and below 40\% for the second derivative of the CCF), and black vertical lines mark the RVs.}
    \label{fig_fig2specAndccf}
\end{figure*}

To address the inherent discreteness of the CCF, we smooth its derivatives by convolving them with a Gaussian kernel. Following the method of \citet{Merle_2017}, we perform a Gaussian smooth and range selection for the CCF and its derivatives. We derive the successive derivatives of CCF, and use Python function \textit{scipy.ndimage.gaussian\_filter1d} \citep{Virtanen_2020} to smooth the derivatives of the CCFs. The $\sigma$ of the Gaussian kernels is initially set as 13 km/s. This is based on our experience calculating CCFs using LAMOST-MRS spectra. A smaller initial value of sigma may lead to more false detections of CCF peaks, while a larger initial value may miss real peaks. We increase the $\sigma$ by 1\,km/s until the number of detected RV components matches the number of valleys in the second derivative or $\sigma$ reaches 100\,km/s. As shown in the left panel of Figure~\ref{fig_fig2specAndccf}, an SB candidate is identified using this method with a final $\sigma$ of 27\,km/s. In the right panel, significant peak blending is demonstrated in the CCF, where the first derivative is insufficient to clearly distinguish the peaks. This shows the importance of calculating higher-order derivatives, as the peaks can be accurately identified using the third derivative.

\begin{figure*}[htbp]
	\centering
	\begin{minipage}{0.49\linewidth}
	    \centering
		\includegraphics[width=0.79\linewidth]{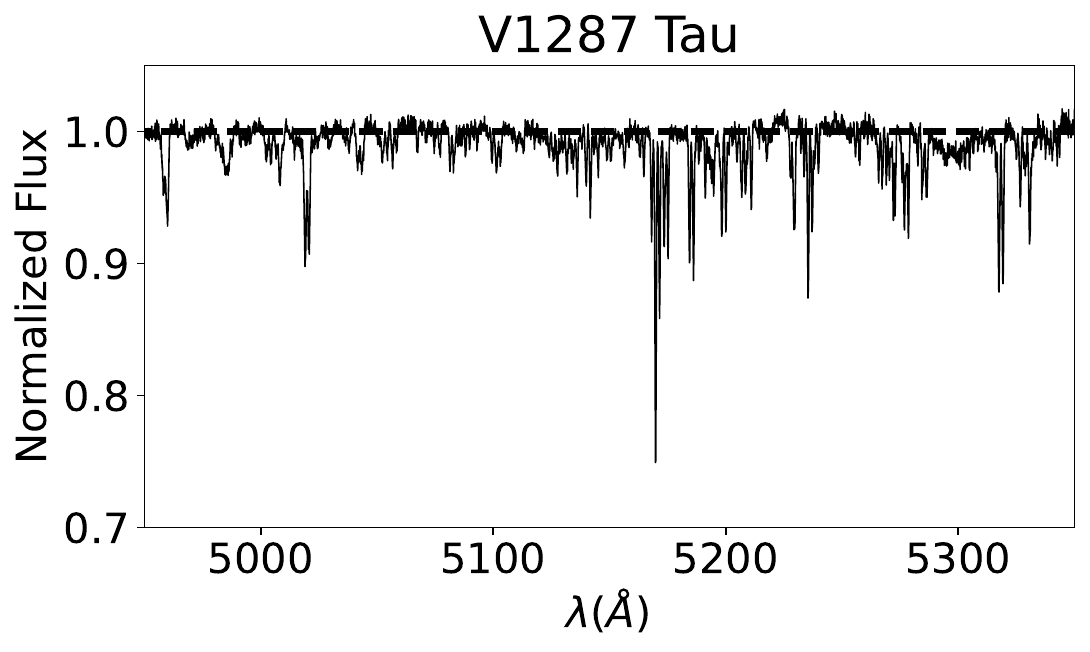}
	\end{minipage}
	\begin{minipage}{0.49\linewidth}
	    \centering
		\includegraphics[width=0.78\linewidth]{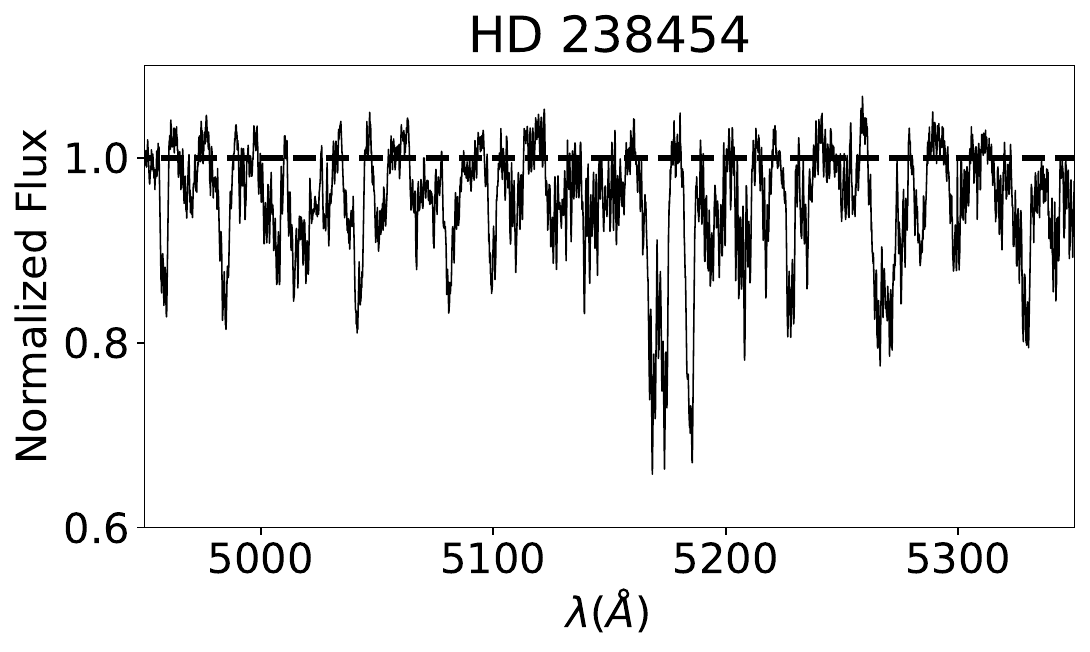}
	\end{minipage}
	\begin{minipage}{0.49\linewidth}
	    \centering
		\includegraphics[width=0.85\linewidth]{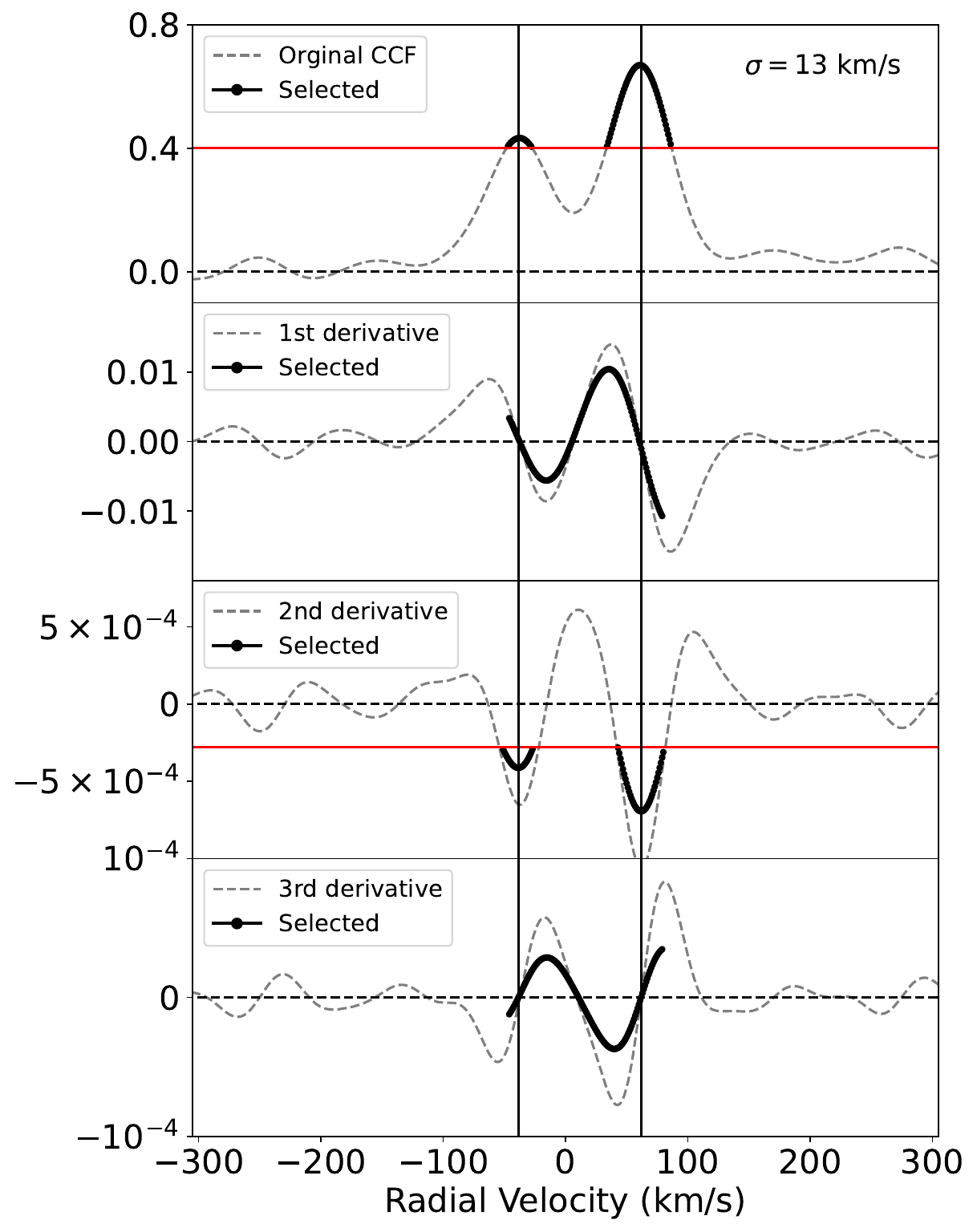}
	\end{minipage}
	\begin{minipage}{0.49\linewidth}
	    \centering
		\includegraphics[width=0.85\linewidth]{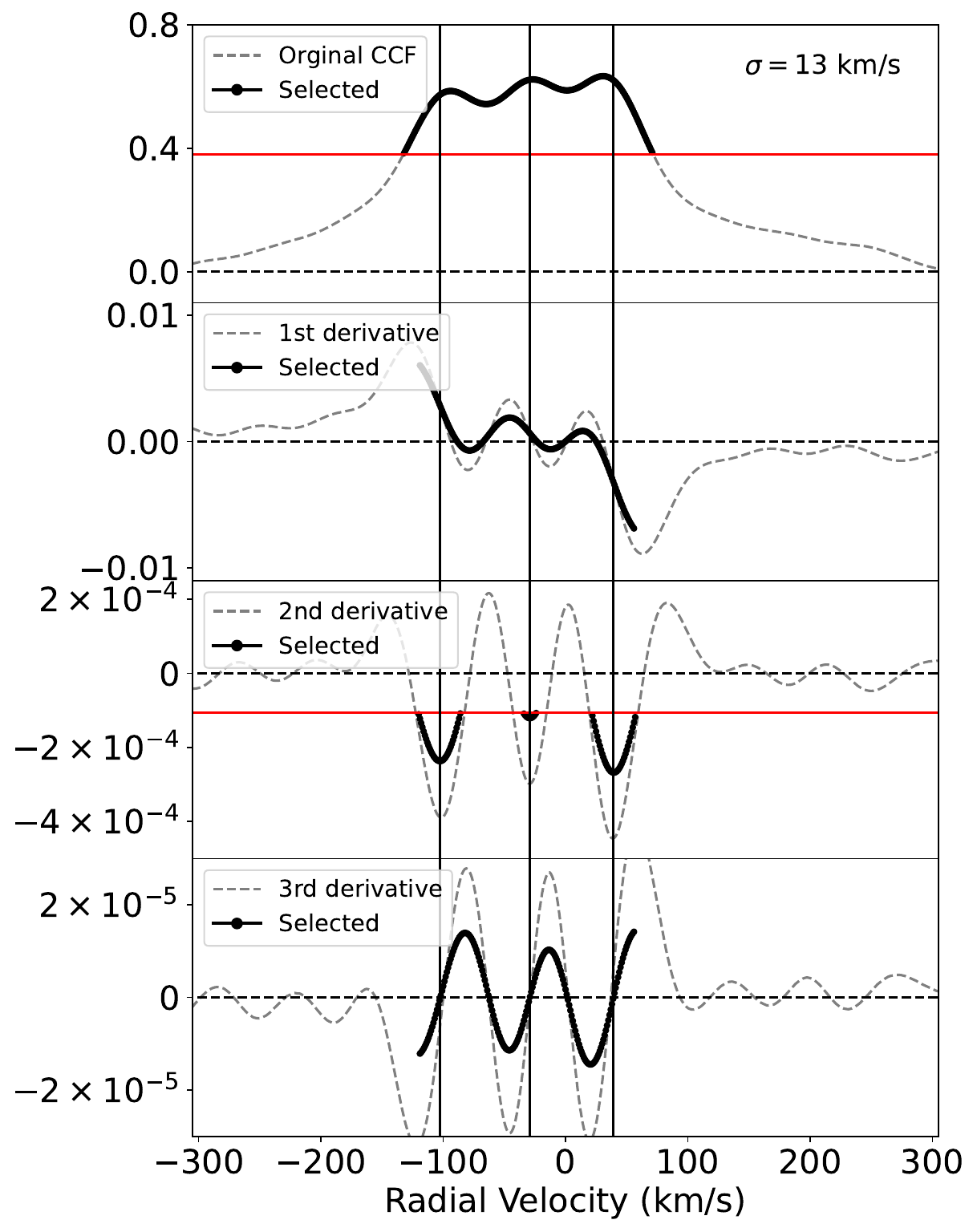}
	\end{minipage}
    \caption{The normalized spectra, CCFs and derivatives of SB2 (V1287 Tau) and SB3 candidates (HD 238454) selected from LAMOST-MRS. The Black solid lines, gray dashed lines, red horizontal lines, and black vertical lines are used as in Figure~\ref{fig_fig2specAndccf}.}
    \label{fig_fig3specAndccf}
\end{figure*}

We select the RV ranges with CCF values higher than 60\% and the second derivative lower than 40\% for RV component detection. The parameters of Gaussian smooth and range selection are based on the experience gained during the detection process for the LAMOST-MRS data. While we acknowledge that lowering these thresholds could potentially increase the detection rate, we prioritized the precision of detected targets over maximizing the number of candidates. This approach is guided by the findings of \citet{Merle_2017}, they presented that a too low threshold can lead to the detection of unrealistic velocity ranges. As shown in Figure~\ref{fig_fig2specAndccf}, selected range of smoothed CCFs and derivatives are drawn with black solid lines and the original CCFs are illustrated in gray dashed lines. The thresholds are drawn in red horizontal lines and the RVs are indicated in black vertical lines. Additionally, the Figure~\ref{fig_fig3specAndccf} show the success of this method in identifying known SB2 candidates, and its applicability for searching SB3 candidates.

\begin{figure*}[ht]
	\centering
	\begin{minipage}{0.3\linewidth}
	    \centering
		\includegraphics[width=0.98\linewidth]{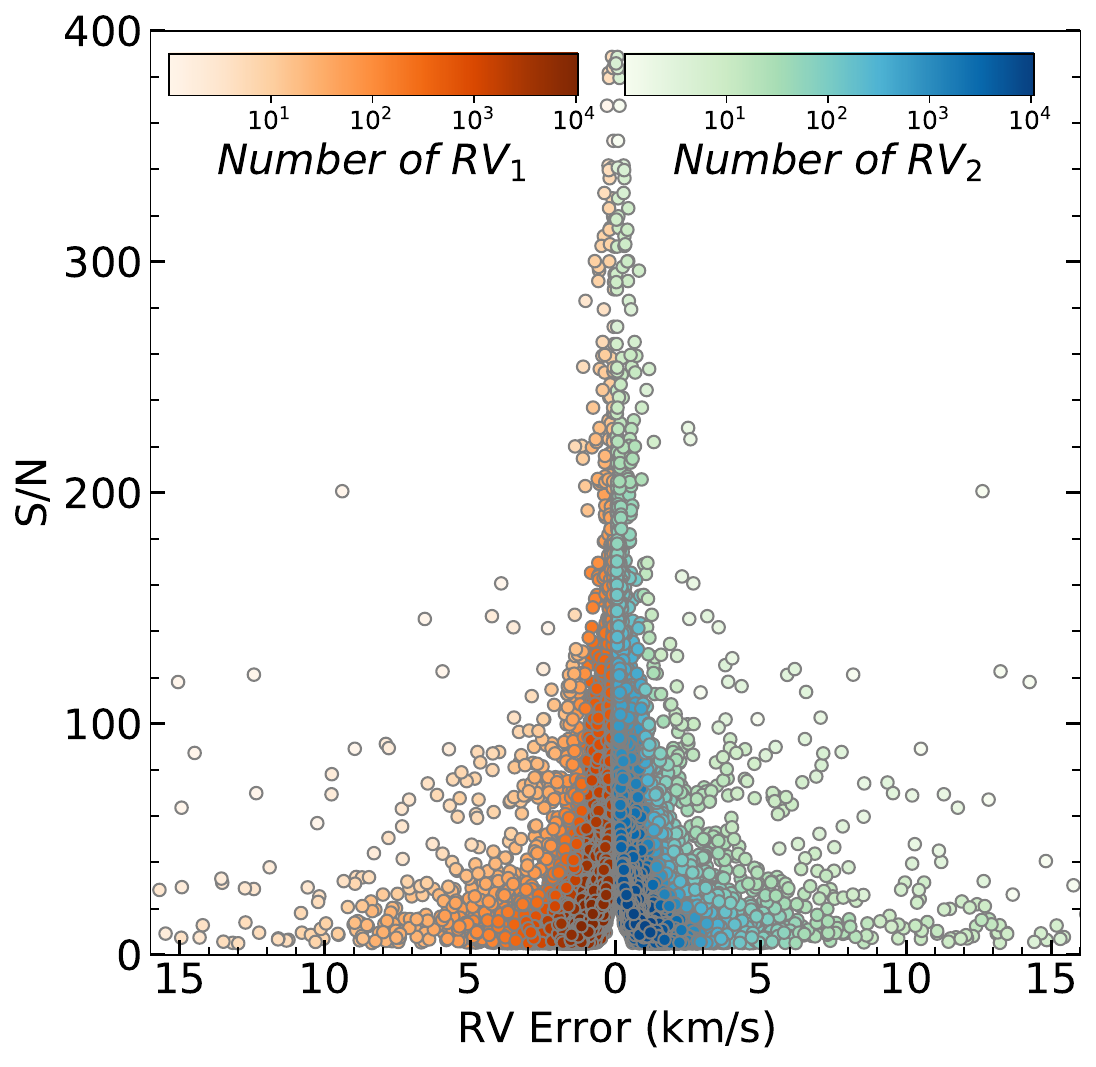}
	\end{minipage}
	\begin{minipage}{0.68\linewidth}
	    \centering
		\includegraphics[width=0.86\linewidth]{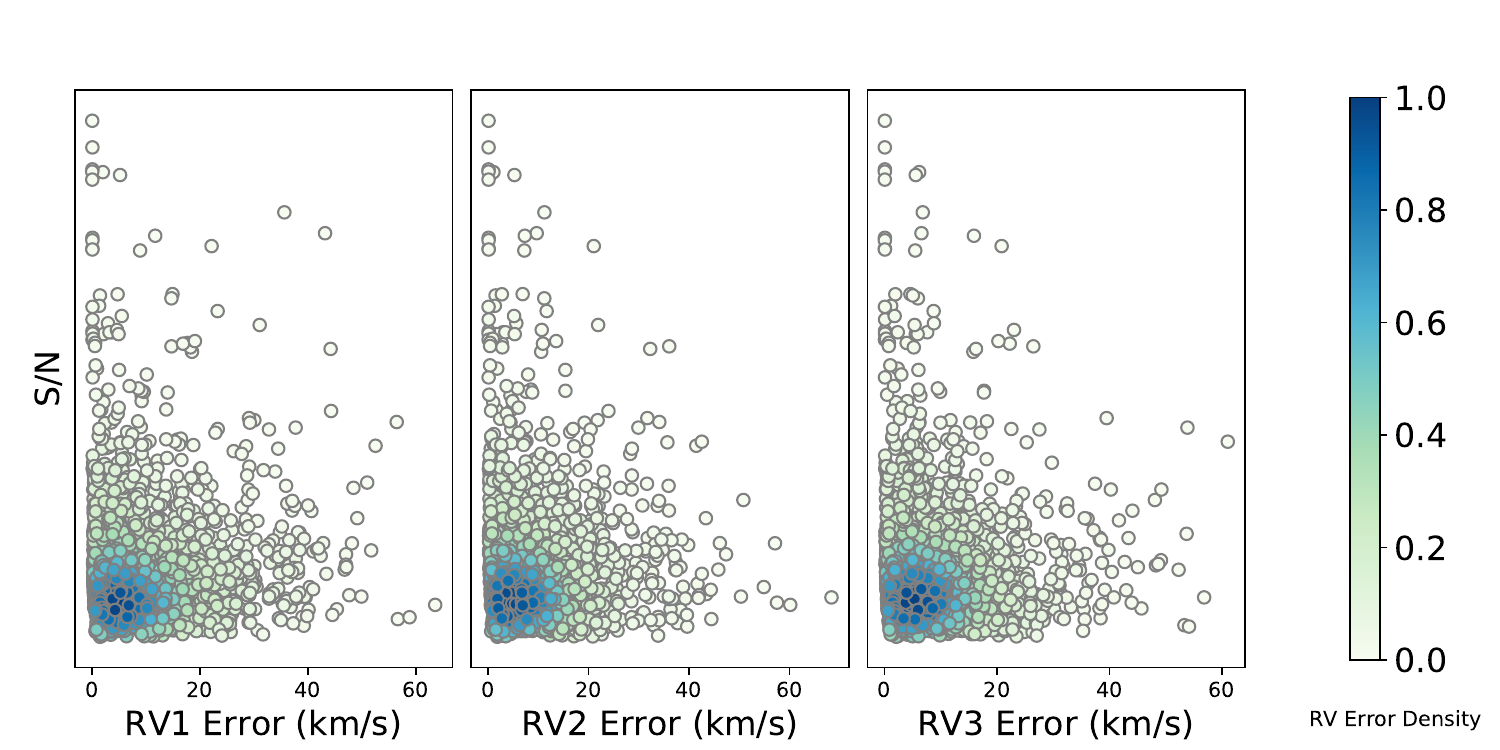}
	\end{minipage}
    \caption{The distribution of RV error versus S/N is illustrated, with the left plot corresponding to the error of RV$_1$ and RV$_2$ of SB2 classification and the right plot corresponding to the RV$_1$, RV$_2$ and RV$_1$ in SB3 classification. The color gradient shows the number density of RV error.}
    \label{fig_fig4rvErrorDistribution}
\end{figure*}

We adopt the Monte Carlo (MC) simulation to estimate the RV uncertainties for each spectrum of double- and triple-line candidates. In this process, we generate 100 simulated spectra by adding random noise to the flux at each wavelength point of observed spectrum. The noise values are randomly generated from a Gaussian distribution, where the mean and variance are defined by the flux and its error in the observed spectrum, respectively \citep{li_double_2021}. This approach allows us to conduct 100 simulations for each spectrum, thereby obtaining the errors and evaluating the reliability and stability of the RV values. The distribution of the uncertainties of RV against S/N is shown in Figure \ref{fig_fig4rvErrorDistribution}.

In the traditional approach, the next step is to manually inspect all targets identified as SB candidates by the CCF calculations for verification. However, using our method and data screening criteria, the number of SB2s identified through the CCF technique could exceed one hundred thousand, making direct visual identification impractical. To enhance efficiency, we introduced the ML method to reclassify the results and expedite the identification process.

\subsection{Machine learning optimization}

We use the CCF data corresponding to each spectrum obtained from the first step for ML classification. ML algorithms tend to perform better in identifying data with clear and distinctive features. Compared to spectral data, the CCFs are considerably smoother and possess more conspicuous features. We utilized deep neural networks (DNNs), a type of multi-layer supervised learning model. DNNs can map input CCFs to output types by leveraging the interconnectedness and weights among multiple layers of neurons to learn complex nonlinear features.  Previous research suggests that increasing the number and diversity of classifiers and adopting multiple methods can enhance classification efficiency and accuracy \citep{Dietterich_2000, Brown_2005, Musehane_2008}, Therefore, we construct multiple classifiers and adopt an ensemble learning strategy to improve the performance of ML.

\subsubsection{Data preparation}

 We use different classification methods based on the characteristics of the data to build four different Classifiers. The final judgment is based on the comprehensive classification results of these Classifiers. The training dataset is composed of four distinct categories, labeled as L0, L1, L2, and L3. L0 represents CCF data with no significant peaks, while L1, L2, and L3 correspond to CCFs of single, double, and triple-line spectra of SBs, respectively. The L0 category indicates either no detectable signal or very weak signals that do not stand out against the noise. It helps the model identify noise data, improving classification performance and reducing false positives. However, the verified number of SB2s and SB3s from LAMOST-MRS observational data is not sufficient for L2 or L3 categories, especially for L3. To address these issues, we use a synthetic spectral library to generate spectra of simulated SBs, and calculate the CCF of these spectra as the training samples.

By applying the ATLAS stellar atmospheric models with the new opacity distribution functions \citep{Castelli_2003}, we calculate the spectra with absolute flux. The wavelength range matches the LAMOST-MRS spectra, spanning $4900 \sim 5400$\,\AA\ with a step of 0.1\,\AA. The grids are in the parameter space with  $T_{\rm eff}$ between 3500\,K and 8000\,K (step 100\,K for 3500 $\sim$ 7500\,K, while 250\,K for 7500 $\sim$ 8000\,K), $\log{g}$ from 0.0\,dex to 5.0\,dex (step 0.25\,dex) and [Fe/H] from $-$4.0 to 0.5\,dex (step 0.2\,dex for $-$4.0 $\sim$ $-$1.0\,dex, while 0.1\,dex for $-$0.1 $\sim$ 0.5\,dex ).

We first generate the spectra of single stars by randomly selecting stellar parameters from the LAMOST-LRS DR9 stellar parameter catalog and interpolate the spectral library in the stellar parameter space to generate the spectra \citep{Xiang_2015}. Then, we select two and three spectra to generate the simulated double- and triple-line spectra, respectively.

\begin{figure*}[htbp]
    \plotone{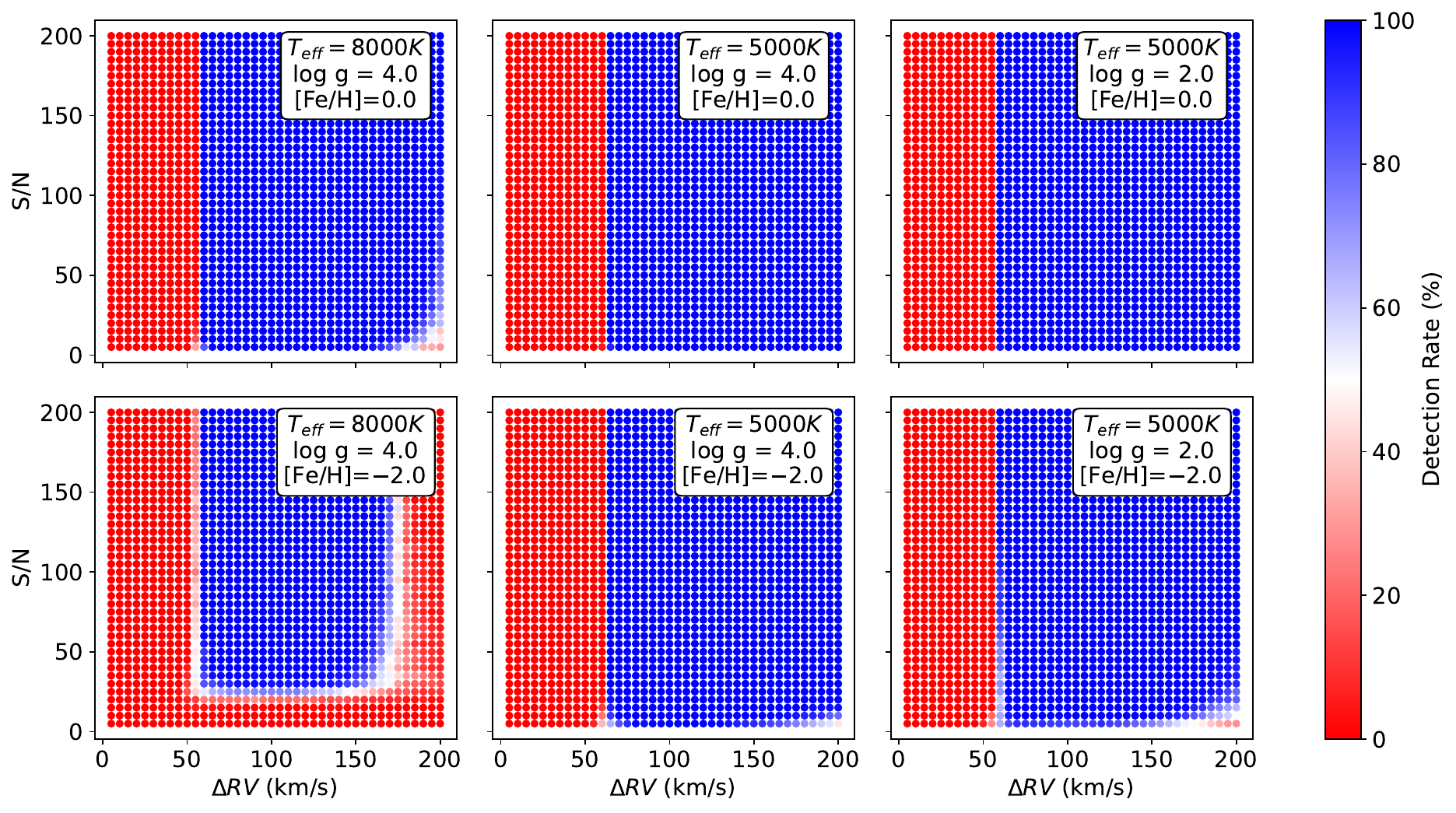}
    \caption{Detection rates of twin stars with different atmospheric parameters, S/Ns and RVs. The minimum S/N used is 5, with a step size of 5.}
    \label{fig_fig5DSR}
\end{figure*}

To generate spectra for SB2 system, we set the RV difference between the two components of the binary is randomly set between 60 and 250\,km/s. For SB3, the relation of the RV of the three components is $RV_1<RV_2<RV_3$. The RV differences between $RV_1$ and $RV_2$, as well as between $RV_2$ and $RV_3$, are randomly set within the range of 60 to 250\,km/s. The lower limit is set due to the resolution of the LAMOST-MRS, which make it difficult to detect stars with RV differences smaller than 60 km/s as shown in Figure~\ref{fig_fig5DSR}. Previous studies have shown that the RV difference between the two components in SBs is rarely greater than 250 km/s \citep{li_double_2021}. Therefore, this upper limit is empirically selected to improve computational efficiency.

\begin{figure}[!ht]
    \plotone{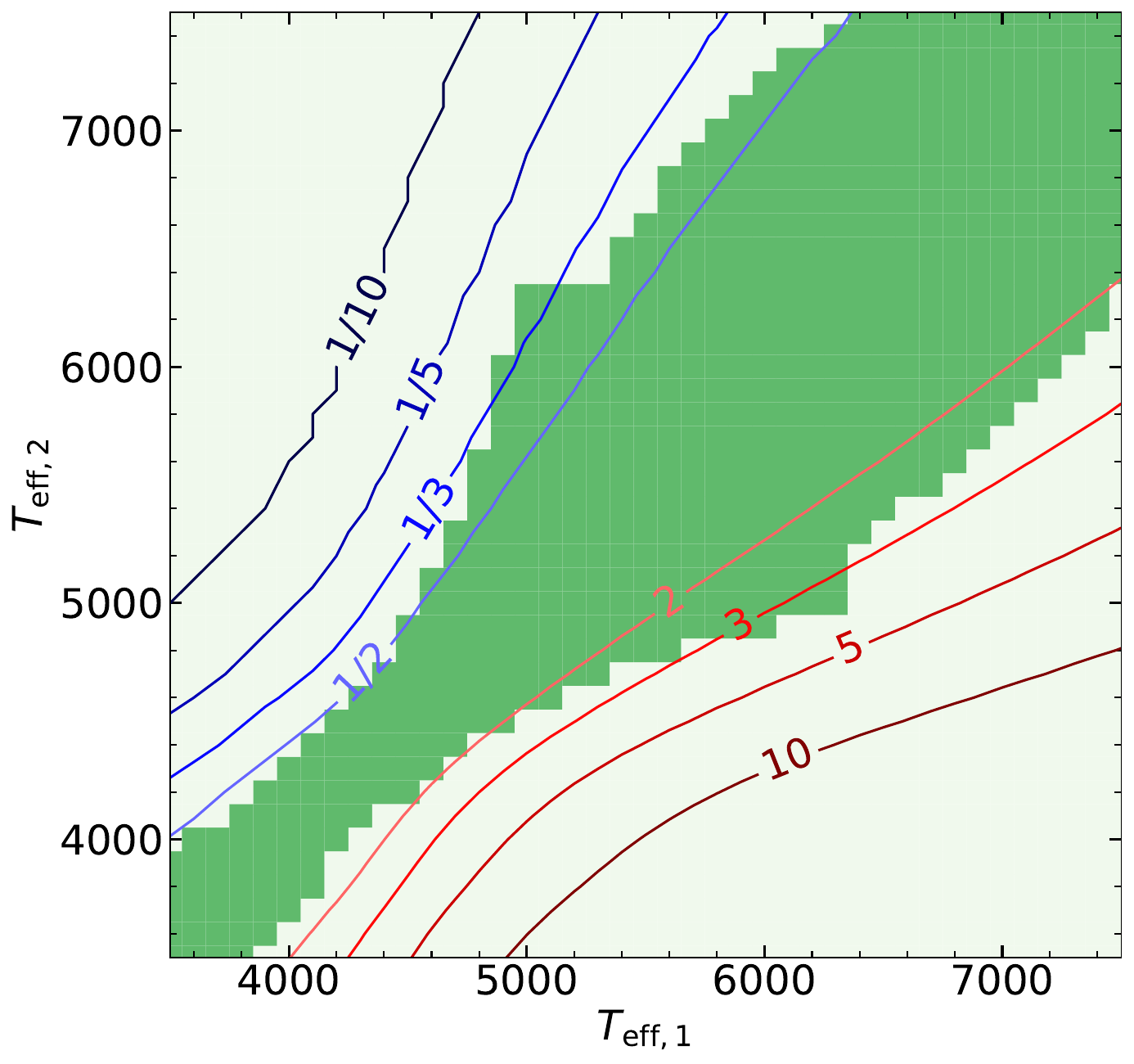}
    \caption{The relationship between successful detection and the spectral flux ratios of the two stars in the binary system. The green blocks indicate successful detection of double-line spectra, while the contour lines depict the spectral flux ratio between the components of the binary stars.}
    \label{fig_fig6FluxRatio}
\end{figure}

The limitation of the CCF technique also needs to be considered, as it determines the range of spectral flux ratios between the two stars in the binary system that the method can resolve.  ``Successful detection'' is defined as using the CCF technique to effectively separate two or three peaks, which represent the simulated components of the SB2 or SB3 systems, respectively.
Figure \ref{fig_fig6FluxRatio} shows the relationship between successful detection and the spectral flux ratios between the two stars in the system. We generate this figure by simulating SB2s, with each component having the same $\log{g} = 4.5$\,dex, [Fe/H] = 0.0\,dex and RV differences are 200km/s. The $T_{\rm eff}$ ranges from 3500\,K to 7500\,K, with a step of 100\,K. Most spectra with a flux ratio between 1/3 and 3 can be detected, and for a higher probability of successful detection, the flux ratio between stars in a binary system should fall within this range. We observe that the flux ratio limit drops below 3 around 5000\,K, which can be explained by the strong and broad MgH molecular lines in K5-type stars, which cause the spectral flux to change more rapidly with $T_{\rm eff}$ around 4500\,K. The increased flux ratio causes greater signal blending, making it more difficult to distinguish the individual spectral features of the stars.

The L0 dataset is selected from CCFs with a maximum CCF value lower than three times the noise level, defined as the standard deviation of the CCF values, to dynamically account for the S/N and spectral quality of different spectra, thereby ensuring a more precise selection. The L1 dataset is formed with single-line spectra. The CCF technique is used on SBs spectra to detect double-line and triple-line features. Only spectra with successful detection are selected to form the L2 and L3 training datasets. Finally, the training dataset includes 1500 samples from the L0 dataset, along with 1500 samples each from the L1, L2, and L3 datasets, for a total of 6000 samples.

\subsubsection{Training and evaluation}

For training DNNs, labeled data are used to adjust the weights and biases of the neural network to minimize the discrepancy between predicted outputs and true labels, thereby improving the prediction accuracy. We adopt Keras with TensorFlow \citep{Abadi_2016} to achieve DNN training.

The neural network architecture includes three fully connected layers implemented on the training set. The first layer consists of 600 neurons and uses Rectified Linear Unit \citep[ReLU,][]{Nair2010RectifiedLU} as the activation function. This is followed by a second layer of 300 neurons, also using ReLU. The final layer comprises $n$ neurons, where $n$ corresponds to the number of output classes, and employs the normalization exponential function softmax as the activation function. When dealing with \textit{n}-class classification problems, the output results of the softmax function, which are the probability values of each class, are obtained by Equation \ref{eq2},

\begin{equation}
\begin{aligned}
\label{eq2}
\text{softmax}(x_i) = \frac{e^{x_i}}{\sum_{j=1}^ne^{x_j}},\quad i=1,2,\cdots,n
\end{aligned}
\end{equation}

\noindent here, \textit{x} is an \textit{n}-dimensional vector composed logits output by the model for different categories, where \textit{n} corresponds to the number of output classes, taking values of 2, 3, or 4 depending on the classification model used by each classifier. After applying softmax function, these logits are converted into normalized probabilities (scores), where each value is between 0 and 1, and the total sums to 1, representing the model's confidence in each category.

The model training adopts the Adaptive Moment Estimation \citep[Adam,][] {Kingma_2014} as optimizer and uses the Sparse Categorical Cross Entropy function shown in Equation \ref{eq3} as the loss function,

\begin{equation}
\begin{aligned}
\label{eq3}
LOSS = -\frac{1}{N} \sum_{i=1}^{N} \log(P_{y_{j}})
\end{aligned}
\end{equation}

\noindent where $N$ is the number of samples in the batch. $y_{i}$ is the true class label (an integer index), $P_{y_{i}}$ is the predicted score for the true class $y_{i}$ from the softmax function output of the model. A test set is used during the training process to evaluate the performance of the model.

\begin{deluxetable*}{cccccccc}[htbp]
    \tablenum{2}
    \label{tab: 4ClassifierResult}
    \tablecaption{The classification results of all calculated CCFs. The last column represents the number of spectra classified as SB2 or SB3 by all machine learning classifiers. For SB2, normalized probabilities (scores) must be greater than 95\%. For SB3, the selection criterion is set at 99\%.}
    \tablehead{
        \colhead{RV calculation classification} & \colhead{Classifier} & \colhead{L0} & \colhead{L1} & \colhead{L2}  & \colhead{L3} & \colhead{Selected results}
    }
    \startdata
        \hline
        \multirow{3}*{SB2 (118,274)} & C1 & 52,661 & 181 & 37,913 & 27,519 & \multirow{3}*{27,233 ($P>95\%$)}\\
        & C2 & 44,219 & 1157 & 38,290 & 34,608 & &\\
        & C3 & 57,436 & 208 & 60,630 & - & &\\
        \hline
        \multirow{3}*{SB3 (49,847)}  & C1 & 22,445 & 6 & 549 & 20,519 & \multirow{3}*{11,904 ($P>99\%$)  }\\
        & C2 & 20,758 & 2 & 676 & 22,083 && \\
        & C4 & 23,365 & 44 & - & 20,110 &&\\
        \hline
    \enddata
\end{deluxetable*}

Based on different classification methods of training data, we constructed four classifiers, named C1 to C4. All four types of data samples (labeled L1 to L3) are directly used to train C1, so the output of this classifier consists of four categories. C2 is a two-step classifier. Initially, it combines samples labeled as L2 and L3 into one category called ``LN", which, along with L0 and L1 samples, undergoes three-class classification training. Then, it uses L2 and L3 data for binary classification. When using the C2 classifier, the data are first divided into three categories, then those classified as LN are reclassified into either L2 or L3. Thus, the final output of C2 is also four categories. C3 uses only L0, L1, and L2 data for three-class classification training, yielding three classification results. Finally, C4 uses only L0, L1, and L3 data for three-class classification training. To maintain data balance across different training categories, each category has 1500 samples. We can use classifiers C1, C2, and C3 to identify SB2 samples from the CCFs obtained with the traditional method, corresponding to CCFs with double peaks. Similarly, we can use C1, C2, and C4 to identify SB3 samples, corresponding to CCFs with three peaks.

We conduct a 10-fold cross-validation process for each classifier, meaning the entire dataset is evenly divided into 10 subsets. In each iteration, one subset is selected as the test dataset, while the remaining nine are combined to form the training dataset. This ensures that each subset serves as the test set once, providing a comprehensive evaluation of our model's performance. This approach evaluates the flexibility and reliability of our model by testing it with different data sets.

After each training iteration, we calculate the values of accuracy, precision, recall, and F1 score for the overall ML method, as well as the precision, recall, and F1 score for L2 and L3 with test dataset for evaluation. We use the \textit{sklearn.metrics} package \citep{Pedregosa_2011} to calculate these metrics for each classifier. A detailed description and explanation of the evaluation metrics are provided in Appendix\,\ref{App: performance_evaluation}. The evaluation results are very promising, as all values are greater than 99\%. This indicates that our classifiers perform very well on the simulated data.

\subsection{Classification criteria}

The final output layer of our DNN models uses the softmax function, which outputs normalized probabilities for each category, with values ranging between 0 and 1. For instance, classifier C1 provides the normalized probabilities that the input CCF data belongs to categories L0, L1, L2, or L3. Typically, the data is assigned to the category with the highest probability. However, to avoid misclassification and improve classification precision, we focus only on data where the normalized probability of L2 or L3 is significantly higher than that of the other categories. We empirically set this threshold at 95\% for L2 selection and 99\% for L3 selection. Additionally, we use the majority voting method from ensemble learning approaches \citep{Dietterich_2000} to integrate the outputs of all four classifiers to obtain the final result. 

Therefore, to classify a spectrum as SB2, its CCF must be categorized as L2 in classifiers C1, C2, and C3, with normalized probabilities exceeding 95\% in each. For SB3, we set the threshold at 99\% for L3 classification in C1, C2, and C4, with reasons discussed in Section \ref{section: Discussion}. Ultimately, as recorded in Table \ref{tab: 4ClassifierResult}, a total of 27,233 CCFs labeled as L2 and 11,904 CCFs labeled as L3 were selected.

\section{Result}

\begin{figure*}[!ht]
	\centering
	\begin{minipage}{0.49\linewidth}
	    \centering
		\includegraphics[width=0.85\linewidth]{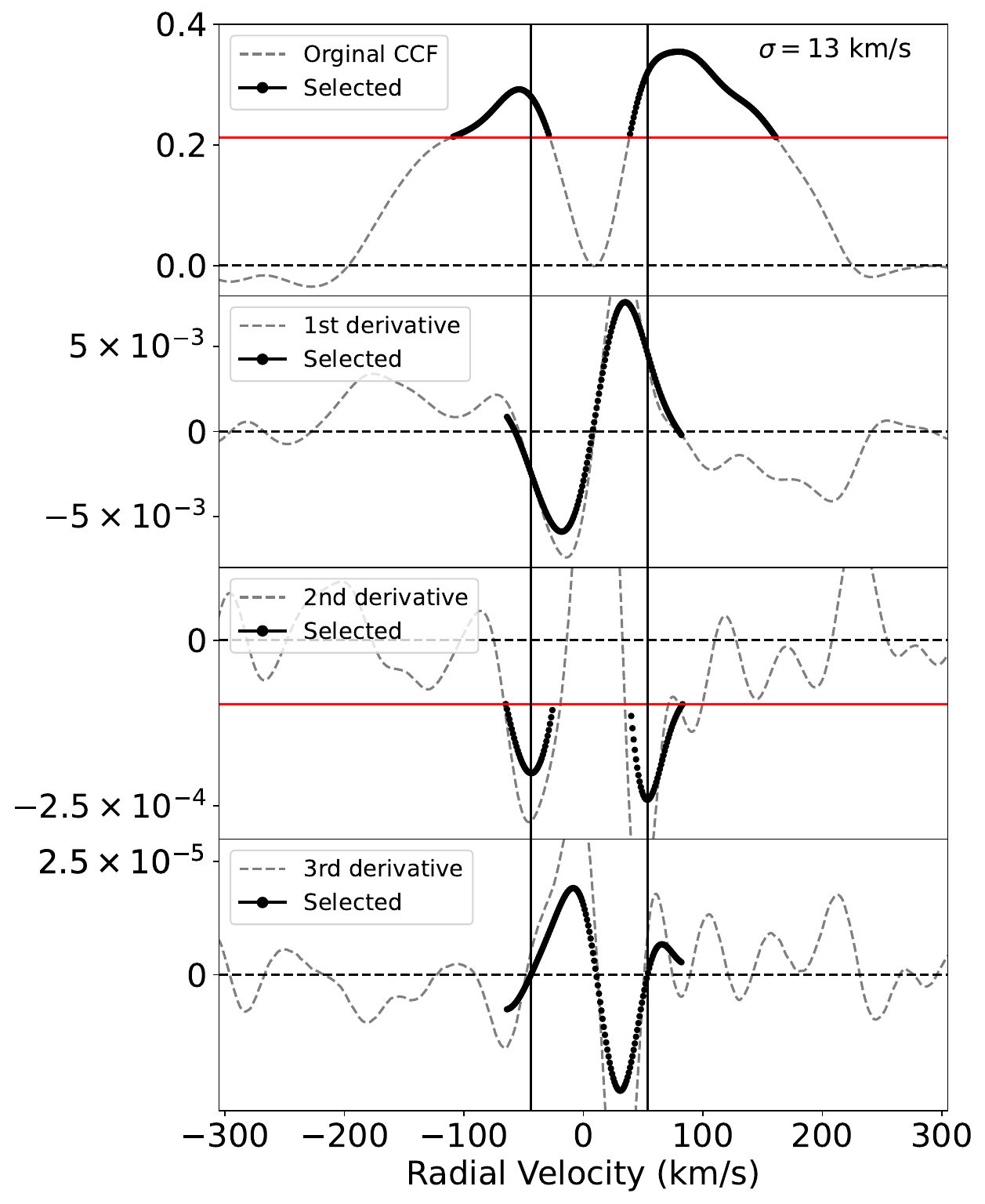}
	\end{minipage}
	\begin{minipage}{0.49\linewidth}
	    \centering
		\includegraphics[width=0.85\linewidth]{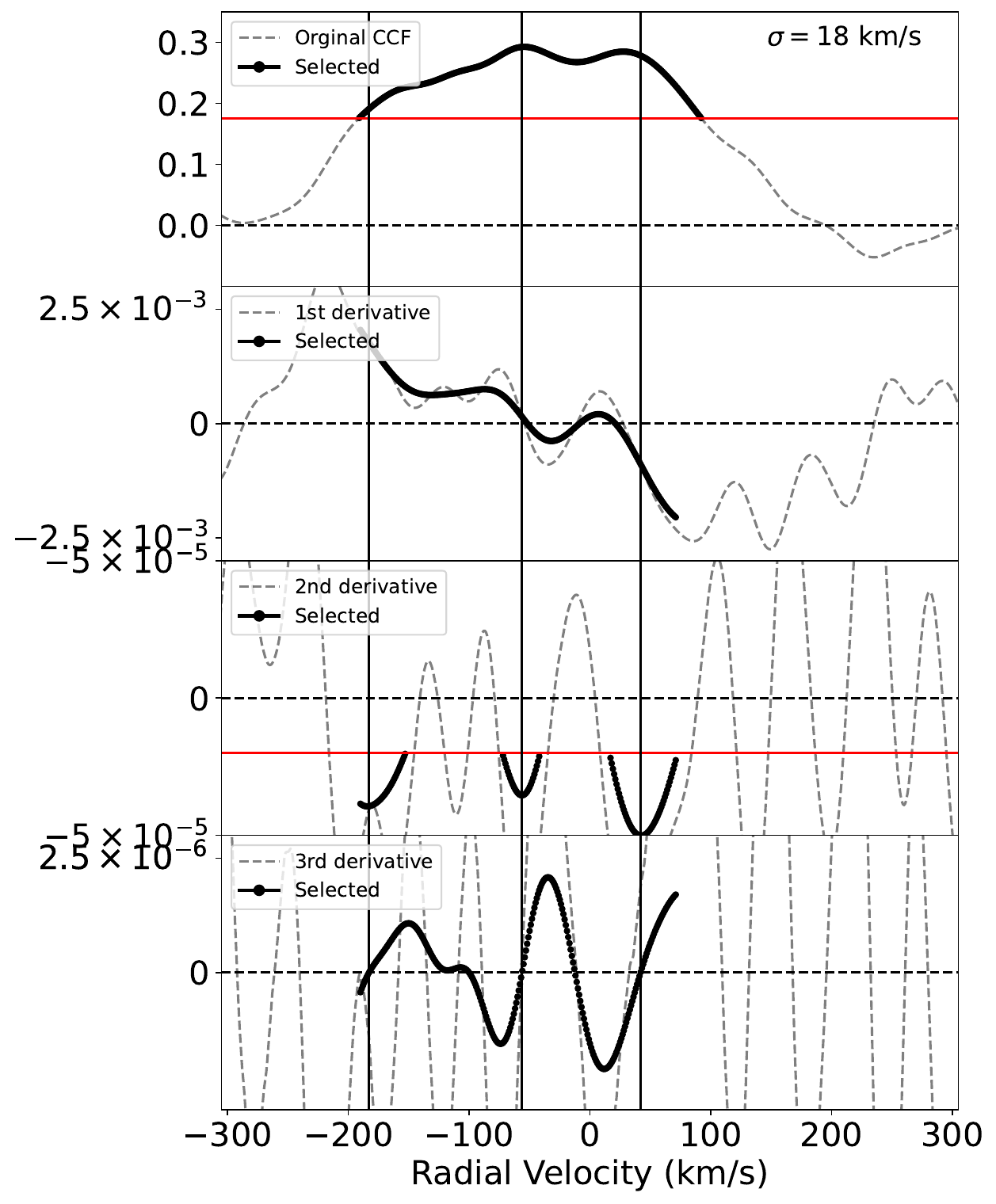}
	\end{minipage}
    \caption{Two examples of cases rejected by visual inspection, including the CCF and its derivatives. The left shows a double-line case identified by the CCF technique and ensemble learning but rejected based on the visual inspection criterion, while the right shows a similar triple-line case.}
    \label{fig_fig7eyeReject}
\end{figure*}

Based on the set of 6,565,721 blue arm spectra selected from the LAMOST-MRS DR9 database, we search for SBs. We obtain the RV component number using the conventional CCF technique, leading to the selection of 118,274 double-line and 43,519 triple-line spectra. The CCF data are also calculated to enable subsequent classification with ML classifiers.

After the ML approach, we identify 27,233 SB2 spectra and 11,904 SB3 spectra. The results are then examined visually to verify the precision of this method, resulting in 27,164 double-line spectra and 3124 triple-line spectra being selected. The screening criterion is that the double-line or triple-line signal in the peak area must be significantly stronger than that in the wing part. CCF data that are rejected typically exhibit broader peaks with greater uncertainties. In the case of double lines, clearly asymmetric CCF peaks are also removed, as they may not be caused by binary stars. As shown in Figure~\ref{fig_fig7eyeReject}, the left is a double-line case while the right a triple-line case, both identified by CCF technique and ensemble learning but rejected through visual inspection.

Given that a target may be observed multiple times with multiple spectra, it is possible for it to exhibit different CCF patterns and consequently be classified into different categories. After careful checking, we find 76 candidates have spectra from different observations classified into both double-line and triple-line. These candidates are ultimately classified as an SB3 candidate and removed from the SB2 category.

After detection and classification using the CCF and ML methods, we identify 7096 SB2 and 1903 SB3 candidates, accounting for 0.8\% and 0.2\% of our selection dataset of LAMOST-MRS DR9 , respectively, using the criteria shown in Table \ref{tab:tableCriteria}. We present the results, including RV values and uncertainties of each spectrum, in Table \ref{tab: sb2sb3candidates}. The table includes plan name (planID), spectrograph ID (spID), fiber ID of the target (fiberID), local modified Julian Minute at the start of exposure (LMJM), median S/N of all pixels (S/N) from the LAMOST-MRS data release. The Local Modified Julian Minute (LMJM) is obtained by multiplying the Local Modified Julian Day (LMJD) of the observation start time by 1440. The LMJD refers to the Modified Julian Day at Xinglong Observatory, which operates in the +8 time zone. \textit{Gaia} source ID (Gaia source id) and G magnitude (G) are cross-matched from \textit{Gaia} DR3 \citep{gaia_2023} using coordinates with a radius of 3.3". The radius is determined from the the diameters of the fiber of LAMOST \citep{Cui_2012}. The RVs in the table are ordered by increasing values.

\begin{longrotatetable}
\centering
\movetabledown=0.5in
\begin{deluxetable*}{ccrccrrrccrrr}
    \centering
    \tablenum{3}
    \tabletypesize{\scriptsize}
    \tablecaption{The information of the spectra identified as double- and triple-line, along with their RVs.}
    \tablewidth{0pt}
    \label{tab: sb2sb3candidates}
    \tablehead{
        \colhead{RA(J2000)} & \colhead{DEC(J2000)} & \colhead{\textit{Gaia} source id} & \colhead{G(mag)} & \colhead{Multiplicity}  & \colhead{planID} & \colhead{spID} & \colhead{fiberID} & \colhead{LMJM} & \colhead{S/N} & \colhead{RV$_1$ (km/s)} & \colhead{RV$_2$ (km/s)} & \colhead{RV$_3$ (km/s)}
    }
    \startdata
    0.011929 & 55.573441 & 420512590034739840 & 12.6 & SB2 & HIP11784201 & 8 & 117 & 83604867 & 24.47 & $ -87.13\pm11.07 $ & $ 30.77\pm8.49 $ & - \\
    0.012862 & 58.019821 & 422632272352729984 & 12.1 & SB3 & NT235302N592517N01 & 7 & 118 & 84169112 & 40.75 & $-118.4\pm7.03$ &$ -21.92\pm2.46 $& $64.26\pm5.25$ \\
    0.012862 & 58.019821 & 422632272352729984 & 12.1 & SB3 & NT235302N592517N01 & 7 & 118 & 84169131 & 39.8 & $-94.07\pm10.9$ &$ 2.5\pm2.66 $& $94.65\pm2.88$ \\
    0.087823 & 55.456971 & 420511593602426240 & 9.1 & SB2 & HIP472H382601 & 3 & 185 & 85224695 & 172.86 & $ -46.5\pm3.32 $ & $ 38.41\pm4.72 $ & - \\
    0.090428 & 38.651075 & 2880989157927180800 & 12.9 & SB2 & NT001046N372826M01 & 16 & 6 & 84706218 & 17.14 & $ -112.81\pm0.47 $ & $ -29.7\pm0.78 $ & - \\
    0.090428 & 38.651075 & 2880989157927180800 & 12.9 & SB2 & NT001046N372826M01 & 16 & 6 & 84706241 & 18.8 & $ -114.05\pm0.47 $ & $ -30.64\pm0.6 $ & - \\
    0.090428 & 38.651075 & 2880989157927180800 & 12.9 & SB2 & NT001046N372826M01 & 16 & 6 & 84706265 & 19.4 & $ -112.91\pm0.38 $ & $ -30.87\pm0.74 $ & - \\
    0.10414 & 34.114326 & 2875121652781859200 & 13.0 & SB2 & NT235421N333536M01 & 9 & 180 & 85231820 & 38.18 & $ -65.59\pm0.44 $ & $ 22.11\pm0.33 $ & - \\
    0.10414 & 34.114326 & 2875121652781859200 & 13.0 & SB2 & NT235421N333536M01 & 9 & 180 & 85231844 & 35.85 & $ -66.32\pm0.41 $ & $ 23.44\pm0.35 $ & - \\
    0.146393 & 55.574784 & 420513135485523456 & 12.6 & SB2 & NGC778901 & 8 & 116 & 83647886 & 18.47 & $ -49.62\pm10.99 $ & $ 46.19\pm8.33 $ & - \\
    0.240085 & 40.398735 & 2881983837993175552 & 12.3 & SB2 & HIP11776901 & 6 & 11 & 83604832 & 15.6 & $ -53.43\pm8.0 $ & $ 31.16\pm7.99 $ & - \\
    0.291179 & 58.695535 & 422768646150973696 & 12.2 & SB2 & NGC778901 & 12 & 240 & 83647859 & 15.02 & $ -14.56\pm1.58 $ & $ 65.22\pm1.36 $ & - \\
    0.291179 & 58.695535 & 422768646150973696 & 12.2 & SB2 & NGC778901 & 12 & 240 & 83647873 & 13.84 & $ -15.11\pm2.2 $ & $ 67.53\pm2.1 $ & - \\
    0.291179 & 58.695535 & 422768646150973696 & 12.2 & SB2 & NGC778901 & 12 & 240 & 83647886 & 16.41 & $ -16.21\pm1.64 $ & $ 65.19\pm1.43 $ & - \\
    0.347601 & 41.11409 & 2882225867991453056 & 12.2 & SB2 & HIP11776901 & 6 & 39 & 83609157 & 21.39 & $ -104.36\pm0.66 $ & $ 26.61\pm0.42 $ & - \\
    0.348095 & 62.072807 & 429910974913478272 & 10.1 & SB3 & NGC778801 & 8 & 78 & 83650805 & 117.37 & $-157.33\pm36.57$ &$ -31.58\pm20.73 $& $65.03\pm13.86$ \\
    0.387174 & 63.517781 & 431593395196335488 & 12.8 & SB2 & NGC778801 & 12 & 23 & 83650762 & 15.39 & $ -83.4\pm0.65 $ & $ -6.5\pm1.01 $ & - \\
    0.40717 & 37.245455 & 2880290044627019264 & 12.4 & SB3 & NT001046N372826M01 & 10 & 237 & 84706241 & 47.87 & $-188.89\pm12.42$ &$ -65.57\pm10.11 $& $63.72\pm22.25$ \\
    0.40717 & 37.245455 & 2880290044627019264 & 12.4 & SB3 & NT001046N372826M01 & 10 & 237 & 84706265 & 44.31 & $-187.82\pm17.33$ &$ -34.73\pm30.52 $& $73.13\pm33.9$ \\
    0.459621 & 61.623334 & 429522400634115840 & 10.9 & SB2 & NGC778802 & 6 & 245 & 83650843 & 54.99 & $ -63.65\pm0.66 $ & $ 23.94\pm0.46 $ & - \\
    0.502711 & 62.14576 & 429912211864024832 & 11.5 & SB2 & NGC778802 & 8 & 99 & 83650826 & 40.49 & $ -59.58\pm3.97 $ & $ 26.38\pm3.84 $ & - \\
    0.527144 & 55.92513 & 420890615868217600 & 10.8 & SB2 & NGC778901 & 8 & 129 & 83647886 & 75.73 & $ -67.09\pm0.39 $ & $ -0.12\pm0.44 $ & - \\
    0.564065 & 60.307202 & 429326137806983424 & 12.7 & SB2 & NGC778801 & 7 & 202 & 83650792 & 16.14 & $ -121.21\pm6.22 $ & $ 20.87\pm18.73 $ & - \\
    0.573322 & 36.428159 & 2880099653021400832 & 12.4 & SB3 & TD000246N354855T01 & 15 & 185 & 84704935 & 51.91 & $-79.75\pm7.35$ &$ 11.44\pm2.68 $& $90.9\pm2.88$ \\
    0.57379 & 62.551303 & 430041125294777088 & 12.2 & SB3 & NGC778802 & 13 & 243 & 83650826 & 33.35 & $-100.19\pm12.77$ &$ 2.1\pm15.27 $& $80.54\pm19.52$ \\
    0.590771 & 36.817228 & 2880214040885803264 & 14.8 & SB2 & TD000246N354855T01 & 15 & 24 & 85227526 & 7.48 & $ -53.97\pm2.72 $ & $ 21.99\pm2.23 $ & - \\
    0.60622 & 61.149567 & 429469074319230592 & 12.8 & SB3 & NGC778802 & 6 & 23 & 83650857 & 15.88 & $-98.87\pm5.67$ &$ -21.61\pm7.25 $& $66.09\pm16.79$ \\
    0.615598 & 35.827237 & 2877066104735658752 & 13.7 & SB2 & NT001046N372826M01 & 2 & 85 & 84706218 & 15.8 & $ -135.35\pm0.79 $ & $ -35.23\pm1.07 $ & - \\
    0.615598 & 35.827237 & 2877066104735658752 & 13.7 & SB2 & NT001046N372826M01 & 2 & 85 & 84706241 & 15.12 & $ -134.49\pm0.8 $ & $ -34.49\pm0.98 $ & - \\
    0.615598 & 35.827237 & 2877066104735658752 & 13.7 & SB2 & NT001046N372826M01 & 2 & 85 & 84706265 & 13.64 & $ -132.74\pm0.9 $ & $ -36.23\pm1.12 $ & - \\
    0.664305 & 33.941192 & 2875153676057907456 & 11.1 & SB3 & TD000246N354855T01 & 1 & 155 & 85227572 & 36.08 & $-89.28\pm14.73$ &$ 44.5\pm42.53 $& $128.19\pm38.21$ \\
    0.673242 & 54.984642 & 420460466311205888 & 12.2 & SB2 & NGC778901 & 7 & 59 & 83647873 & 28.19 & $ -161.98\pm1.01 $ & $ 56.83\pm4.2 $ & - \\
    0.673242 & 54.984642 & 420460466311205888 & 12.2 & SB2 & NGC778901 & 7 & 59 & 83647886 & 27.12 & $ -161.8\pm1.16 $ & $ 56.2\pm3.14 $ & - \\
        $\cdots$ & $\cdots$ & $\cdots$ & $\cdots$ & $\cdots$ & $\cdots$ & $\cdots$ & $\cdots$ & $\cdots$ & $\cdots$ & $\cdots$ & $\cdots$ & $\cdots$ \\
    \enddata
    \tablecomments{The table presents the classification outcomes, including RV values and associated errors. The columns are: RA(J2000); DEC(J2000); \textit{Gaia} source ID from \textit{Gaia} DR3; G(mag) is the G magnitude from \textit{Gaia}; Multiplicity, which is the result of the classification; planID is the plan name from LAMOST-MRS; spID is the spectrograph ID from LAMOST-MRS; fiberID is the fiber ID of the target from LAMOST-MRS; LMJM, showing the local modified Julian Minute at the start of exposure; S/N representing the median S/N of all pixels; RV$_1$(km/s), RV$_2$(km/s), and RV$_3$(km/s), which are the RVs calculated from the first step using the conventional CCF technique and arranged in order of their values. The full version of this table with 30,288 rows.}
\end{deluxetable*}
\end{longrotatetable}

\subsection{SB2 and SB3 Candidates}

We cross-match our SB2 and SB3 candidates with other binary catalogs including the Kepler Eclipsing Binary Stars \citep[KEBC,][]{Kirk_2016}, the TESS Eclipsing Binary stars \citep[TESS-EBs,][]{Andrej_2022}, SB in the APOGEE DR16 and DR17 Data \citep{Kounkel_2021}, Gaia DR3 Non-single stars \citep{gaia_2022},  SB candidates from Gaia-ESO Survey \citep{Merle_2017}, FGK binary stars from the GALAH survey \citep{Traven_2020}, the $\rm S_{B^9}$ catalog\citep{Pourbaix_2004}, Wide binaries from Gaia eDR3 \citep{el_badry_2021_4435257} and the Washington Visual Double Star Catalog \citep[WDS,][]{Mason_2001}. As the diameter of the LAMOST fiber is about 3.3", we cross-match catalogs and obtain the common targets by coordinates within a 3.3" radius. A total of 690 common SB2 and 151 common SB3 candidates have been obtained.

We also cross match the candidates obtained by \citet{li_double_2021} using LAMOST-MRS DR7 data and find 1637 SB2 and 58 SB3 candidates identified in both works. Taking into account of all the cross match results, 2121 SB2 and 197 candidates identified in this work have been included in other catalogs or studies to our knowledge (Table \ref{tab:commonSB2SB3}). Therefore, 4975 SB2 and 1706 SB3 candidates are newly found. The proportion of new discoveries in all identified SB2 and SB3 candidates is 70.1\% and 89.6\%, respectively.

\begin{deluxetable*}{ccccccc}[htbp]
    \tablecaption{The list of common SB2 and SB3 candidates with other binary catalogs.}
    \tablenum{4}
    \label{tab:commonSB2SB3}
    \tablehead{
         \colhead{RA(J2000)} & \colhead{DEC(J2000)} & \colhead{Gaia Source ID} & \colhead{G(mag)} & \colhead{Multiplicity} & \colhead{$N_{\rm Exp}$} & \colhead{Comment}}
    \startdata
        0.291179 & 58.695535 & 422768646150973696 & 12.2 & SB2 & 3 & LC\_2021 \\
        0.347601 & 41.114090 & 2882225867991453056 & 12.2 & SB2 & 1 & Gaia WB,LC\_2021 \\
        0.387174 & 63.517781 & 431593395196335488 & 12.8 & SB2 & 1 & LC\_2021 \\
        0.407170 & 37.245455 & 2880290044627019264 & 12.4 & SB3 & 2 & WDS \\
        0.459621 & 61.623334 & 429522400634115840 & 10.9 & SB2 & 1 & Gaia WB \\
        0.527144 & 55.925130 & 420890615868217600 & 10.8 & SB2 & 1 & WDS \\
        0.673242 & 54.984642 & 420460466311205888 & 12.2 & SB2 & 2 & LC\_2021 \\
        0.863206 & 29.936055 & 2861247637104271360 & 11.7 & SB2 & 1 & Gaia WB \\
        1.354247 & 58.314799 & 422698316070242816 & 11.3 & SB2 & 2 & LC\_2021 \\
        1.417710 & 56.414781 & 420974865952459904 & 11.6 & SB2 & 3 & Gaia WB,LC\_2021 \\
        1.513075 & -0.538712 & 2545718964216072320 & 13.7 & SB2 & 3 & LC\_2021 \\
        1.650851 & 57.137139 & 422510368300513792 & 9.6 & SB2 & 3 & LC\_2021 \\
        2.144856 & 58.668384 & 422906772301985408 & 12.5 & SB3 & 1 & Gaia NSS \\
        2.184212 & 37.497105 & 2877637747703176704 & 11.4 & SB2 & 1 & WDS \\
        2.275254 & 61.646542 & 429848096591774208 & 12.4 & SB2 & 1 & LC\_2021 \\
        2.482804 & -0.206435 & 2545836543240450816 & 13.4 & SB2 & 1 & LC\_2021 \\
        2.742505 & 54.741034 & 420218882982724224 & 13.9 & SB2 & 4 & WDS \\
        2.796898 & 57.199411 & 422425740266770048 & 11.3 & SB2 & 3 & LC\_2021 \\
        2.870514 & 56.897016 & 422327918091786752 & 11.8 & SB2 & 3 & LC\_2021 \\
        2.882273 & 57.712906 & 422469097952109568 & 13.5 & SB2 & 5 & LC\_2021 \\
        2.905883 & 59.259216 & 423035316384989824 & 13.3 & SB2 & 5 & LC\_2021 \\
        2.942407 & 58.377895 & 422884717652219776 & 12.9 & SB2 & 3 & Gaia WB,WDS \\
        3.332230 & 2.609186 & 2548317385070710656 & 13.9 & SB2 & 3 & LC\_2021 \\
        3.397440 & 2.621560 & 2548316904034412416 & 14.2 & SB2 & 1 & LC\_2021 \\
        3.495918 & -1.915404 & 2541576382359366784 & 13.8 & SB2 & 3 & LC\_2021 \\
        3.648931 & 53.828180 & 420110443650955904 & 11.1 & SB2 & 4 & WDS \\
        3.931877 & 58.901100 & 422964977698391168 & 10.1 & SB2 & 6 & LC\_2021 \\
        3.943690 & 57.117233 & 422389284583657472 & 14.0 & SB2 & 1 & LC\_2021 \\
        4.059039 & 58.118580 & 422821942408901376 & 10.8 & SB2 & 1 & Gaia WB \\
        4.070429 & 59.209457 & 428978108012689792 & 11.1 & SB3 & 1 & Gaia NSS \\
        4.275290 & 39.547891 & 379956794398809472 & 11.9 & SB2 & 3 & Gaia WB \\
        4.651834 & 2.108786 & 2548075939189064320 & 13.9 & SB2 & 3 & LC\_2021 \\
        5.081090 & 58.854170 & 428204189266719616 & 13.6 & SB2 & 6 & LC\_2021 \\
        5.198236 & 60.060925 & 428383890690176512 & 10.0 & SB2 & 4 & WDS \\
        5.316476 & 60.702454 & 428799368656549120 & 13.3 & SB2 & 2 & LC\_2021 \\
        5.519838 & 57.614777 & 421996449694222720 & 13.6 & SB2 & 1 & Gaia WB \\
        5.868530 & 24.272019 & 2801215161221470720 & 13.9 & SB3 & 1 & Gaia NSS \\
        6.140866 & 59.600082 & 428307612078727936 & 13.4 & SB2 & 4 & LC\_2021 \\
        6.256972 & 23.492209 & 2800927638930707712 & 12.2 & SB2 & 3 & Gaia WB \\
        6.355069 & 59.116074 & 428267308105062528 & 11.1 & SB2 & 6 & Gaia NSS,LC\_2021,WDS \\
        6.508332 & 32.691329 & 2862935284373463168 & 13.7 & SB2 & 3 & Gaia WB \\
        $\cdots$ & $\cdots$ & $\cdots$ & $\cdots$ & $\cdots$ & $\cdots$ & $\cdots$ \\
        \hline
    \enddata
    \tablecomments{The list of common SB2 and SB3 candidates with other binary catalogs. The columns are: RA (J2000); DEC (J2000); \textit{Gaia} source ID from \textit{Gaia} DR3; G(mag) is the G magnitude from \textit{Gaia}; Multiplicity, which is the result of the classification; $N_{\rm Exp}$ is the number of exposures of this candidate in LAMOST-MRS, Comment listing the catalogs in which this candidate is included. In the table, abbreviations represent various star catalogs: KEBC (Kepler Eclipsing Binary Stars), TESS (TESS Eclipsing Binary Stars), APOGEE (SB in APOGEE DR16 and DR17 Data), Gaia NSS (Gaia DR3 Non-single stars), Gaia-ESO (SB candidates from Gaia-ESO Survey), GALAH (FGK binary stars from GALAH survey), SB9 (SB9 catalog), Gaia WB (Wide binaries from Gaia eDR3), WDS (Washington Visual Double Star Catalog), and LC\_2021 \citep{li_double_2021}. The full version of this table with 2318 rows. }
\end{deluxetable*}

\subsection{Statistical analysis}

\begin{figure}[!t]
    \plotone{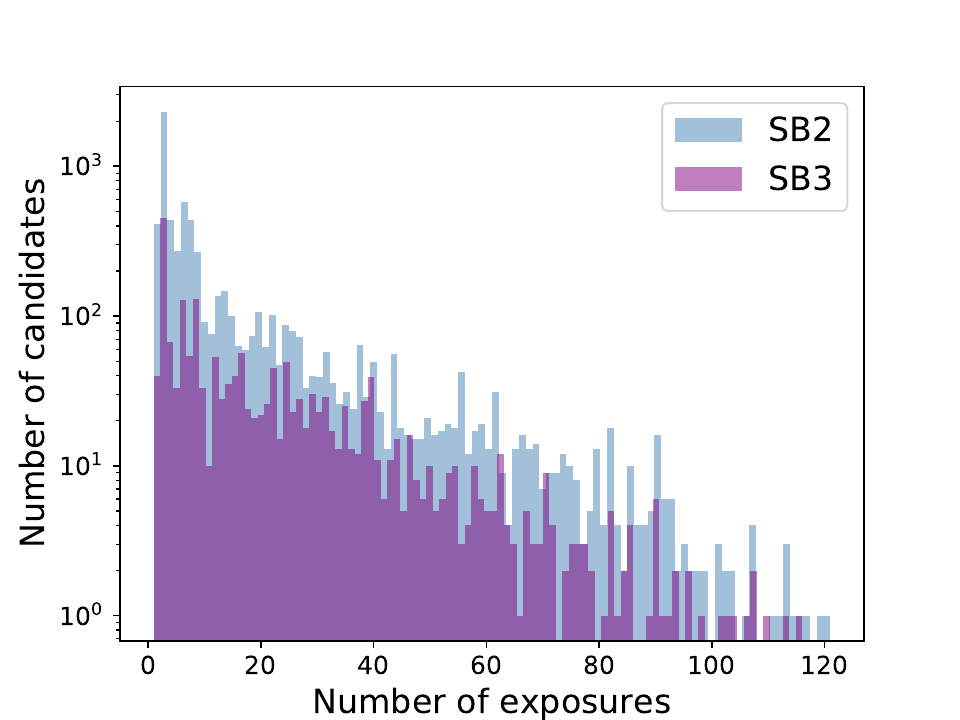}
    \caption{Number of detected SB candidates versus number of exposures (Blue arm). The number of candidates is in a logarithmic scale.}
    \label{fig_fig8expoNo}
\end{figure}

We perform a statistical analysis on all the 7096 SB2 and 1903 SB3 candidates identified. For each candidate, we determine the number of spectra (blue arm) in the LAMOST-MRS DR9 database that can yield RV values using our methods. The distribution of SBs as a function of the number of exposures is illustrated in Figure \ref{fig_fig8expoNo}. Among these candidates, 108 SB2 and 7 SB3 candidates have single exposures, with the highest number of exposure times being 121 for SB2 and 116 for SB3. There are 3650 SB2 and 1312 SB3 candidates have exposures over 6 and the orbital parameters for this sample of SB2 candidates could be obtained.

\begin{figure}[!t]
    \plotone{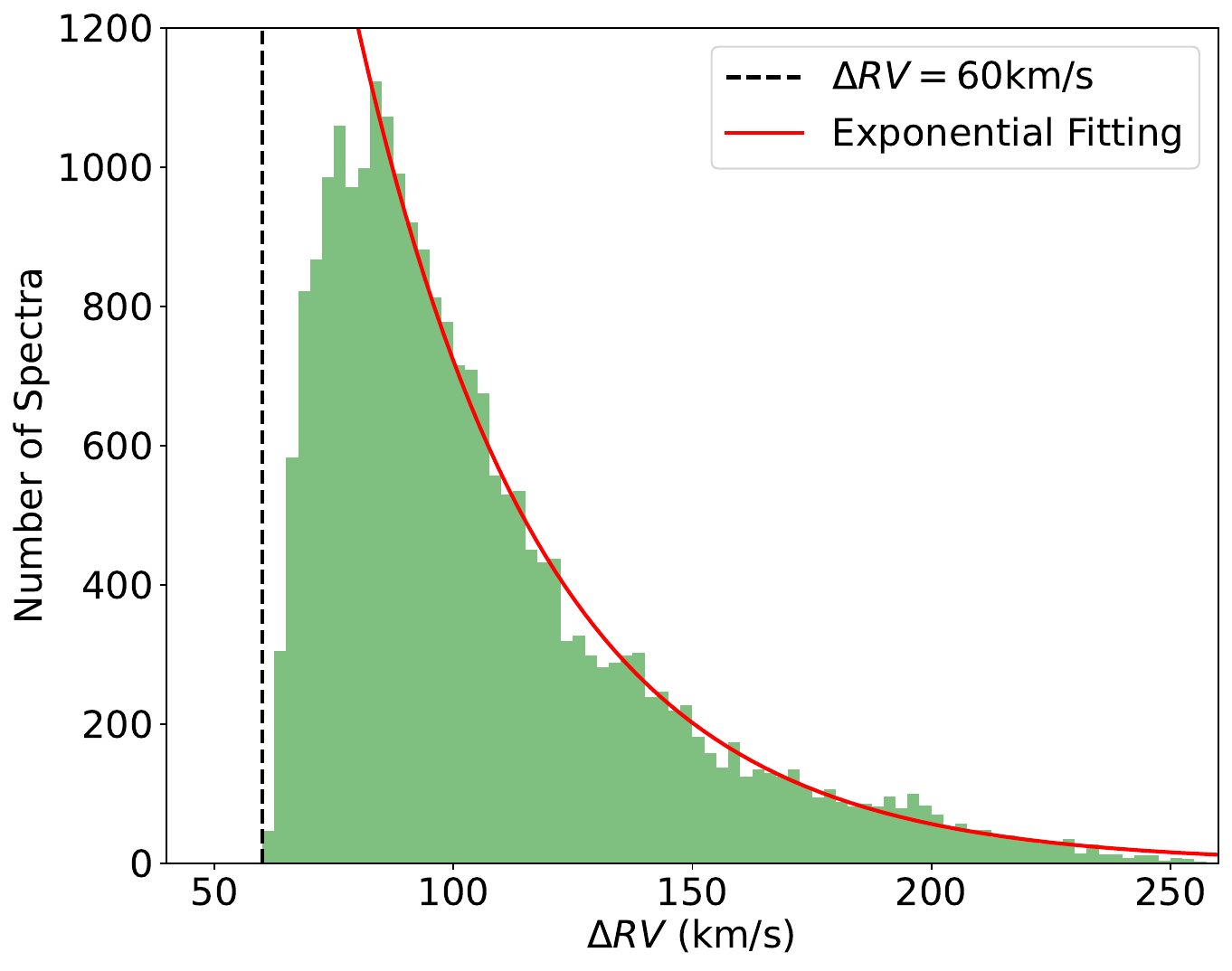}
    \caption{Distribution of RV differences ($\Delta RV$) of all observed SB2 candidates in LAMOST-MRS DR9. The vertical dashed black line indicates the detection limit of $\Delta RV$ for SB2 in LAMOST-MRS spectra, which is about 60 km/s. The red line represents the exponential fitting curve. We applied an exponential function to fit the distribution of $\Delta RV > 80$ km/s.}
    \label{fig_fig9RVveri}
\end{figure}

\begin{figure*}[!ht]
	\centering
	\begin{minipage}{0.49\linewidth}
	    \centering
		\includegraphics[width=0.95\linewidth]{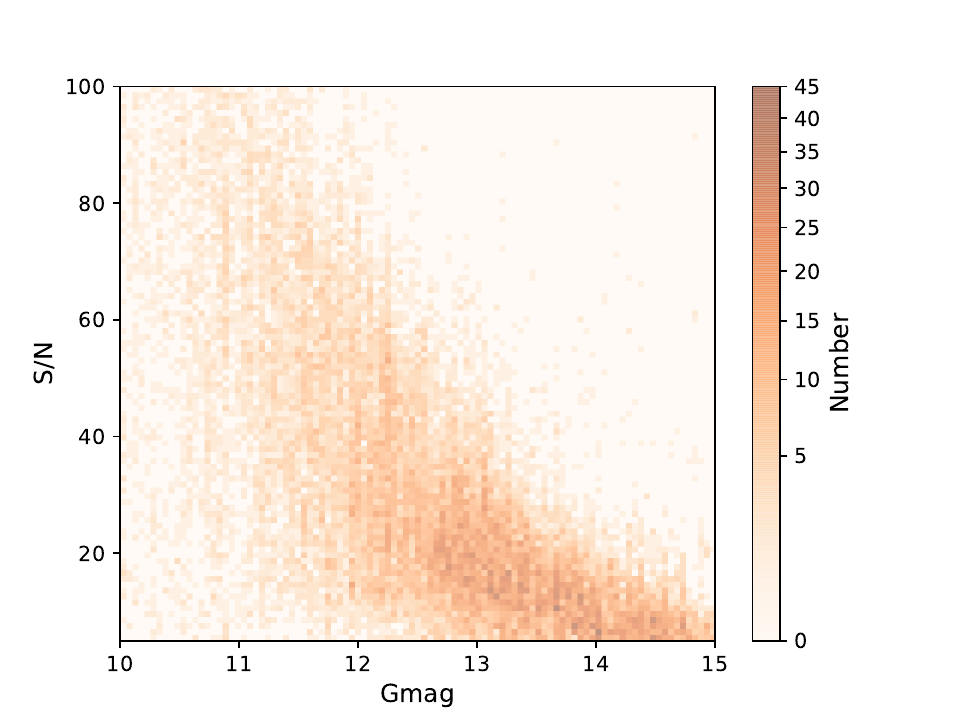}
        \textbf{double-line}
	\end{minipage}
	\begin{minipage}{0.49\linewidth}
	    \centering
		\includegraphics[width=0.95\linewidth]{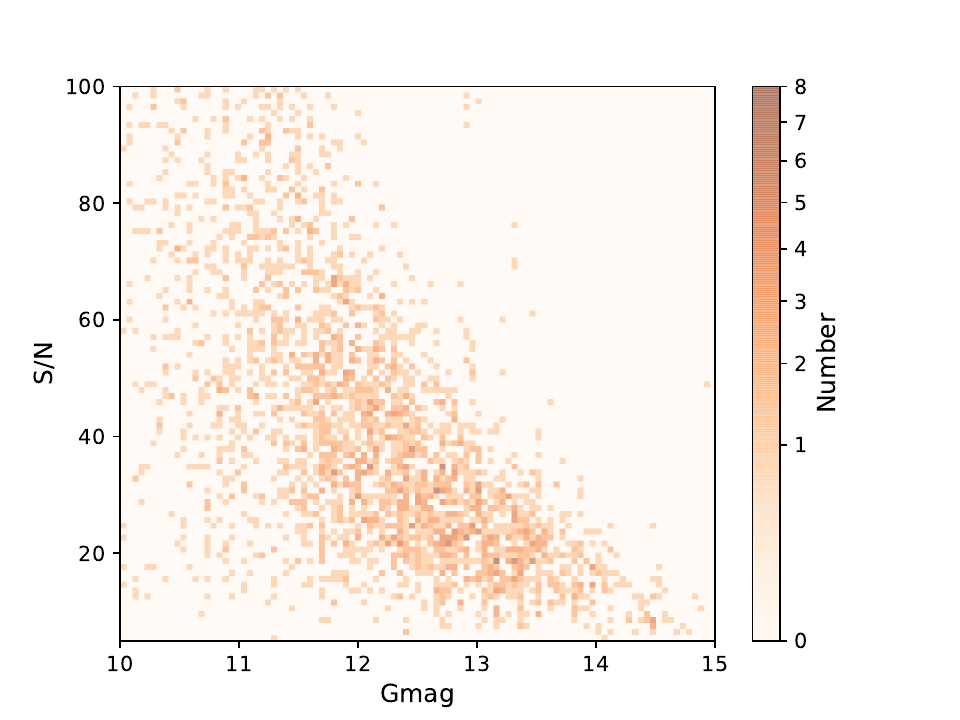}
        \textbf{triple-line}
	\end{minipage}
    \caption{Left panel: The S/N versus G magnitude distribution of 27,164 manually confirmed double-line spectra in LAMOST-MRS DR9. Right panel: The S/N versus G magnitude distribution of 3124 manually confirmed triple-line spectra in LAMOST-MRS DR9.}
    \label{fig_fig10snrGmag}
\end{figure*}

In Figure \ref{fig_fig9RVveri}, the distribution of RV differences for each spectrum of the SB2 candidates are illustrated, and the RV differences $\Delta RV$ are basically larger than 60 km/s, which is related to the resolving power of LAMOST-MRS. Also, due to the resolution limitations, detecting all SBs with $\Delta RV < 80$ km/s is challenging. The distribution of $\Delta RV > 80$ km/s follows an exponential pattern. Figure \ref{fig_fig10snrGmag} shows the distribution of S/N versus G magnitude of SB2 and SB3, and its distribution has the same trend as Figure \ref{fig_fig1SNR}.

\section{Discussion} \label{section: Discussion}

Combining the traditional CCF technique with the ML methods, our human-AI hybrid process significantly improved the precision and efficiency of searching and classification SB candidates from the LAMOST-MRS spectra. In the conventional CCF technique step, we identify 118,274 spectra as double-line candidates. This number significantly exceeds the results from \citet{li_double_2021} and manual examination of the result at this scale is impractical. After ML process, the number of spectral classified as double-line candidates have been reduced from 118,274 to 27,233, and 27,164 are confirmed visually. Assuming the spectra confirmed by human validation as ground truth, the ML method improves the precision from 23.0\% to 99.7\%. As for triple-line spectra, after ML process, the number has been reduced from 43,519 to 11,904, and 3124 spectra are identified visually. The ML method improves the precision from 7.2\% to 26.3\%.

From our work, the identified 7096 SB2 and 1903 SB3 candidates account for 1.0\% of the LAMOST-MRS DR9 data we selected for SB searching. This fraction is slightly lower than the fraction reported by \citet{li_double_2021} using LAMOST-MRS DR7 data. This is mainly because we set the spectral screening criteria of S/N to greater than 5 and processed a significantly larger number of spectra initially.

The reliability and interpretability of ML methods applied in astronomical research remains a subject of debate. This skepticism is largely due to the perceived ``black box'' nature of these methods \citep{Smith_2023}. Nonetheless, human-AI collaboration, which integrates ML approaches with traditional methods based on rigorous mathematical rules, has the potential to enhance the practicality and accuracy of ML methods. This topic is gaining increasing attention \citep{Djorgovski_2022}.

\begin{figure*}[htbp]
	\centering
	\begin{minipage}{0.49\linewidth}
	    \centering
		\includegraphics[width=0.85\linewidth]{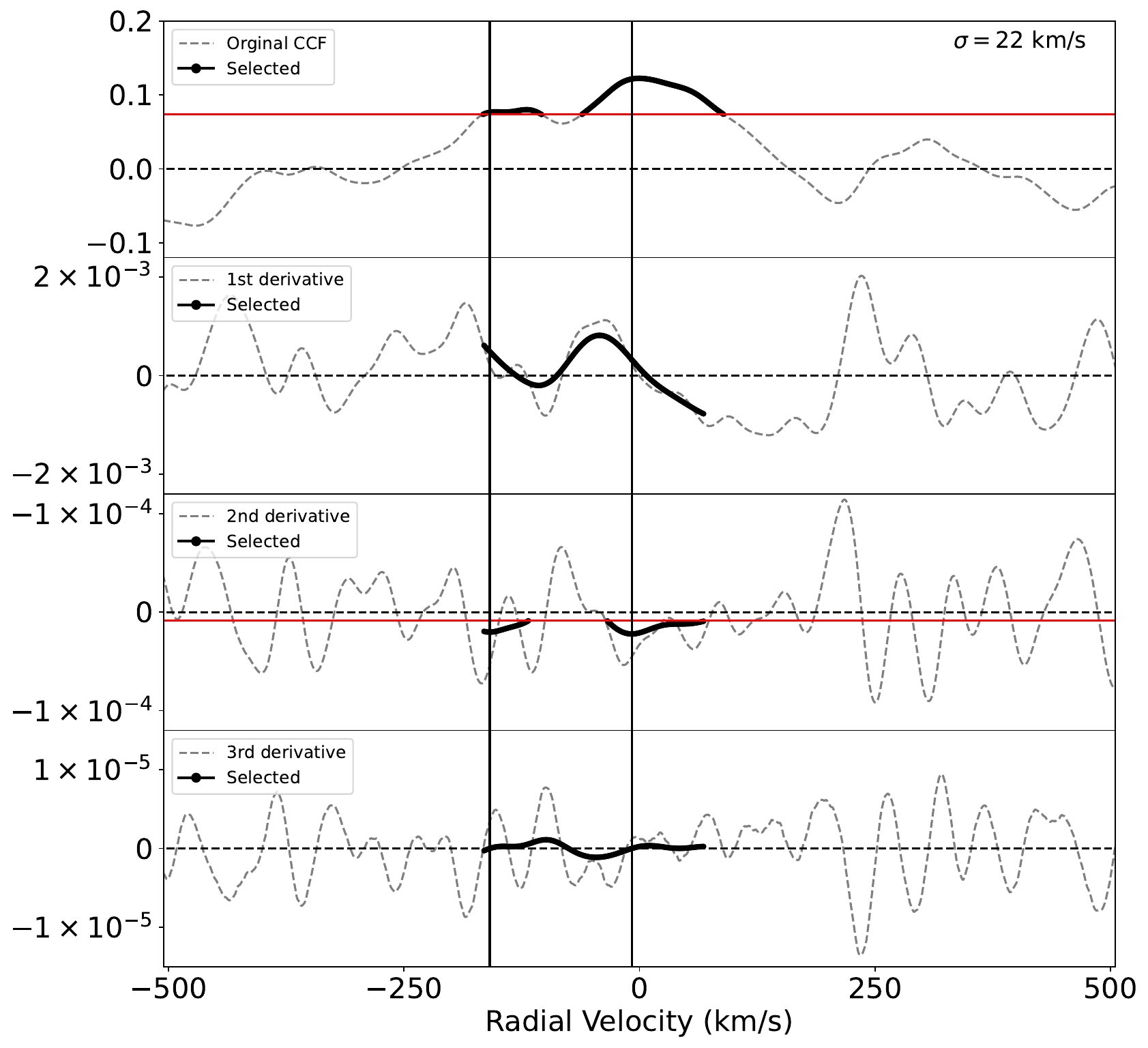}
	\end{minipage}
	\begin{minipage}{0.49\linewidth}
	    \centering
		\includegraphics[width=0.85\linewidth]{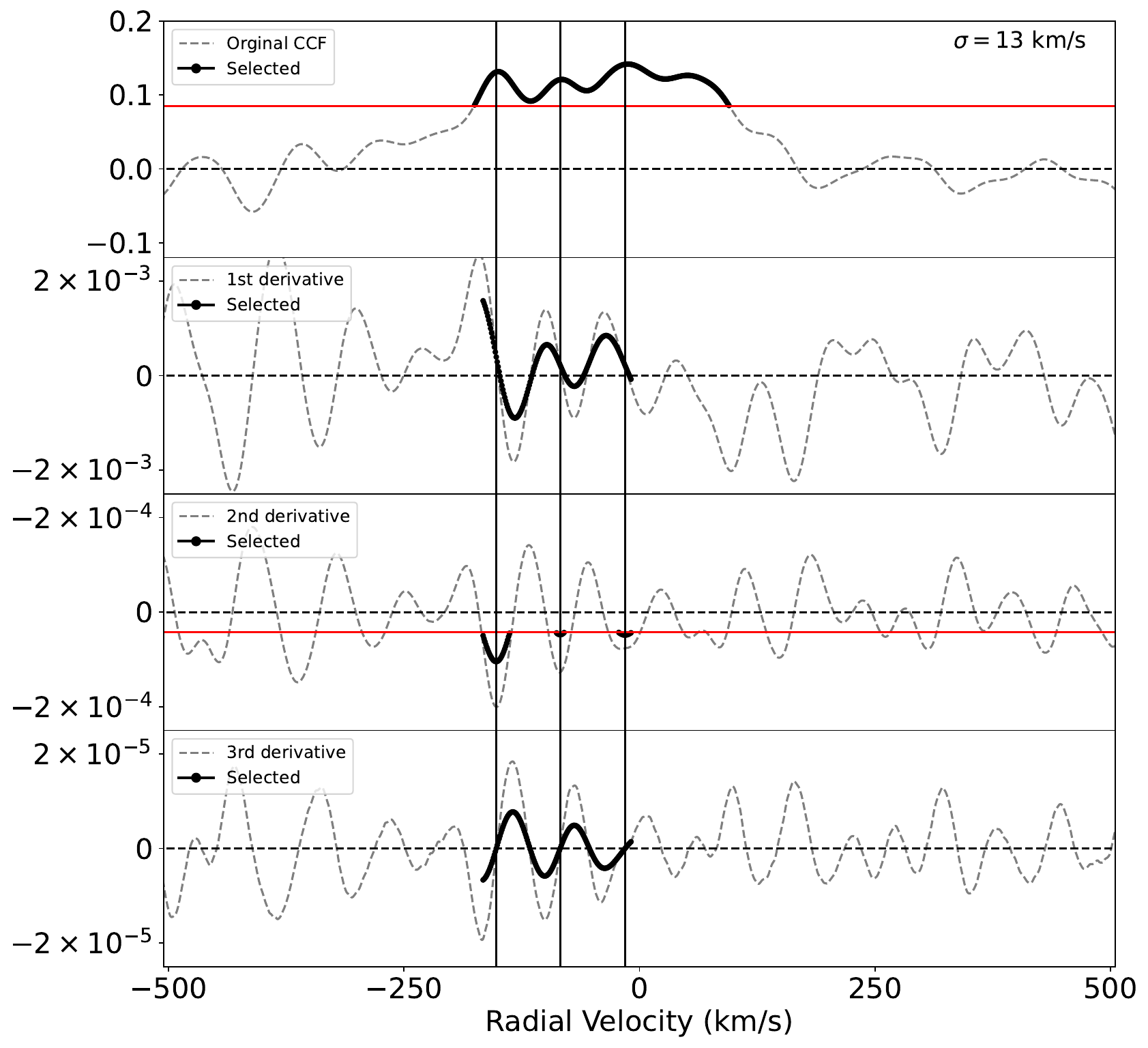}
	\end{minipage}
    \caption{Two CCF data identified as double-line (left) and triple-line (right) by the CCF technique, but filtered out by the ensemble learning classifier.}
    \label{fig_fig1MLReject}
\end{figure*}

We can evaluate the precision of the ensemble learning method and each DNN classifier using classification result in Table \ref{tab: 4ClassifierResult} and the spectra identified visually.  The precision of L2 is 99.7\% for ensemble learning and are 71.6\%, 70.1\% and 44.8\% for classifier C1, C2 and C3, respectively. The precision of L3 is 19.4\% for ensemble learning result and 11.3\%, 10.5\% and 11.5\%. As shown in the Figure~\ref{fig_fig1MLReject}, the two CCFs are respectively identified as double-line and triple-line by the CCF technique, but they are filtered out by the ensemble learning classifier. The ML classifiers likely filtered out the CCFs with the peak signals that are not significantly greater than the noise. This indicates that the model is effectively prioritizing precision and successfully eliminating data that could otherwise introduce false positives.

For our SBs search efforts, we focus more on the precision of ML in identifying L2 and L3 classes. For L2, we can see that our ensemble learning method achieves very high overall precision, nearly matching the results obtained during model training with theoretical spectra. This is because our ML approach is specifically designed to prioritize precision, aiming to reduce the need for manual inspection and enhance overall efficiency. We apply high selection thresholds across multiple classifiers and use their intersection for final selection. As a result, the precision in detecting SB2 candidates approaches 100\%. However, true SB2 candidates may still be included in the spectra that are filtered out, highlighting the need for further improvement in accuracy.

For the L3 category, both precision and accuracy are still require further improvements. This may be due to the fact that our training data does not fully reflect the distribution of real L3 samples, as well as the random selection of RV differences, which does not fully account for the hierarchical relationship of three components in SB3 system. Moreover, the impact of training samples on classification results cannot be ignored. In our study, we performed multiple training sessions using training sets formed from different data types: pure observational data, a combination of observational and theoretical data, pure theoretical data, and theoretical data with physical constraints. When applying these trained models to the spectra to be classified, some spectra, especially those of SB3, could only be identified by a single model but were missed by others. The influence of different training data on the results is particularly significant for SB3 spectra.

\section{Conclusion and Prospects}

We employ a hybrid human-AI approach, consisting of three steps: traditional CCF analysis, ML methods, and human-eye inspection, to detect SB2 and SB3 candidates using the blue-arm spectra from the LAMOST-MRS DR9 data with S/N higher than 5. Initially, number of components and RV values are obtained through CCF calculations producing a preliminary classification. Subsequently, ML methods are introduced to further screen the data and improve the precision of classification. We employ the DNN to build four classifiers and adopt ensemble learning approach. With the human inspection processes, we finally identify a total of 27,164 double-line spectra and 3124 triple-line spectra from the LAMOST-MRS DR9 data, corresponding to 7096 SB2 and 1903 SB3 candidates, respectively. Notably, 70.1\% of the SB2 candidates and 89.6\% of the SB3 candidates are newly identified.

By combining conventional CCF and ML methods, the precision and efficiency of identifying multi-system candidates are significantly improved. Specifically, the precision for SB2 candidates has improved from 23.0\% to 99.7\%. This indicates that the combined method can effectively automate the search for binary stars in LAMOST-MRS data, saving a significant amount of time on visual inspection. However, the results for SB3 candidates are not satisfactory. Our future plans involve further optimizing the process to enhance classification efficiency and precision, particularly for SB3 candidates.

To improve classification performance, we plan to add penalty parameters and apply multi-channel training while increasing the number of samples. Additionally, we are examining the spectra misclassified by the CCF technique (as shown in Figure~\ref{fig_fig1MLReject}) and those misclassified by the ensemble learning method but rejected through visual inspection (as shown in Figure~\ref{fig_fig7eyeReject}), attempting to identify their patterns.

This work is supported by the National Key R\&D Program of China(2022YFF0711500), National Natural Science Foundation of China (NSFC)(12090040, 12090044, 12373110, 12273077, 12103070, 11833006), the 14th Five-year Informatization Plan of Chinese Academy of Sciences (CAS-WX2021SF-0204). Data resources are supported by China National Astronomical Data Center (NADC), CAS Astronomical Data Center and Chinese Virtual Observatory (China-VO). This work is supported by Astronomical Big Data Joint Research Center, co-founded by National Astronomical Observatories, Chinese Academy of Sciences and Alibaba Cloud. Guoshoujing Telescope (the Large Sky Area Multi-Object Fiber Spectroscopic Telescope LAMOST) is a National Major Scientific Project built by the Chinese Academy of Sciences. Funding for the project has been provided by the National Development and Reform Commission. LAMOST is operated and managed by the National Astronomical Observatories, Chinese Academy of Sciences. This work also supported by the Strategic Priority Research Program of the Chinese Academy of SciencesGrant No.XDB0550100.

\bibliographystyle{aasjournal}
\bibliography{main}

\appendix
\setcounter{figure}{0}

\section{Model Performance Evaluation} \label{App: performance_evaluation}
\renewcommand{\thefigure}{A\arabic{figure}}

We employed a standard method for model evaluation, utilizing performance metrics accuracy, precision, recall, and F1 score, as delineated in Equations \ref{eq9} through \ref{eq12}. Where, true positive (TP) represents the positive sample been successfully identified as positive, true negative (TN) is the negative sample correctly identified as negative, false positive (FP) is the negative sample mistakenly classified as positive, and false negative (FN) is the positive sample categorized as negative by the model. 

\begin{equation}
\begin{aligned}
\label{eq9}
Accuracy = \frac{TP + TN}{TP + TN + FP + FN}
\end{aligned}
\end{equation}
\begin{equation}
\begin{aligned}
\label{eq10}
Precision = \frac{TP}{TP + FP}
\end{aligned}
\end{equation}
\begin{equation}
\begin{aligned}
\label{eq11}
Recall = \frac{TP}{TP + FN}
\end{aligned}
\end{equation}
\begin{equation}
\begin{aligned}
\label{eq12}
F1\,\,\,score = \frac{2 * (Precision * Recall)}{(Precision + Recall)}
\end{aligned}
\end{equation}

Using L2 as an example, when considering L2 as the positive sample, it indicates that our analysis focuses on samples labeled as L2:

\begin{itemize}
  \item A true positive (TP) occurs when a sample labeled as L2 is correctly identified as L2 by the model.
  \item A true negative (TN) happens when a sample not labeled as L2 is correctly recognized as not L2.
  \item A false positive (FP) is when a sample not labeled as L2 is incorrectly classified as L2.
  \item A false negative (FN) occurs when a sample labeled as L2 is mistakenly identified as not L2 by the model.
\end{itemize}

In machine learning (ML), accuracy, precision, recall, and F1 score are key indicators for evaluating the performance of classification models. Each metric measures the performance of the model from a different perspective. A high accuracy indicates that the model has a high rate of correct predictions overall or for a specific category. A high precision means the model rarely misclassifies negative samples as positive. A high recall indicates the model captures most of the positive samples, with few misses. The F1 score is the harmonic mean of precision and recall, providing a balance between them. A high F1 score signifies good performance in both aspects, making it especially suitable for evaluating imbalanced datasets.

\end{document}